\documentclass[a4paper,12pt]{article}
\usepackage[utf8]{inputenc}
\usepackage{graphicx}
\usepackage{graphics,epstopdf}
\usepackage{amssymb,amsmath}
\usepackage{booktabs} 
\usepackage{xcolor}
\usepackage{cite}
\usepackage{caption}
\usepackage{subcaption}
\usepackage{geometry} 

\usepackage[a4paper=true,pagebackref=false]{hyperref} 
\hypersetup{colorlinks = true, linkcolor = red, anchorcolor = red, citecolor = blue, filecolor = red, pagecolor = red, urlcolor = red}

\title{Cosmological Model in $f(R, \mathcal{G})$ Gravity : Stability Analysis and Observational Constraints from  DESI DR2}

\author{ Shivani Sharma\footnote{\textbf{corresponding author}: Shivani Sharma, shivani1506@bhu.ac.in}, R. Chaubey\footnote{yahoo\underline{ }raghav@rediffmail.com, rchaubey@bhu.ac.in} \\
Centre for Interdisciplinary Mathematical Sciences \\Institute of Science, Banaras Hindu University\\ Varanasi, Pin 221005, India}

\begin{document}

\maketitle

\begin{abstract}
In this article, we examine the dynamical system of the Decoupled  Power-law $f(R,\mathcal{G})$ gravity model. This $f(R,G)$ model framework is composed of interactions between dark matter and scalar field through the linear coupling term. The key objective of the present study is to describe the cosmological viability of the modified gravity theory formulated with gravity $ f(R, \mathcal{G}) $.  We transform the cosmological equations into an autonomous system of ordinary differential equations by suitable transformation of variables. The decoupled power-law $ f(R,\mathcal{G})$ model governed by $ f(R, \mathcal{G}) = \alpha R^m + \beta \mathcal{G}^n $ has been investigated in detail to characterize the stability properties of the critical points of the autonomous system. The model may explain the late-time accelerating universe expansion corresponding to the attractor in the model. Depending on the effective equation of state parameter values corresponding to the critical points, we study the observational viability of the model using low-redshift observational data, such as observational Hubble data. Furthermore, we investigate the effects of parameters using the effective equation of the state parameter and the statefinder diagnostics. We further investigate the observational viability of the model by constraining its parameters through Markov Chain Monte Carlo (MCMC) analysis of the combined cosmic chronometer, Pantheon+SH0ES, CMB, and DESI DR2 BAO data. The constrained model predicts $H_0 = 69.71 \pm 0.61~\mathrm{km\,s^{-1}\,Mpc^{-1}}$, $q_0=-0.530$ with a transition redshift $z_t=0.646$, and a quintessence-like effective equation of state. The cosmographic parameters and cosmic age are also consistent with $\Lambda$CDM expectations.
\end{abstract}

\textbf{Keywords}:  FRW Cosmological model, Dynamical systems, Dark energy, Coupling parameter,  $f(R,\mathcal{G})$ gravity.

\section{Introduction}\label{sec1}

One of the most fascinating problems in modern cosmology is to figure out how the universe formed and evolved with time. The General Theory of Relativity (GTR) revolutionized the understanding of cosmic dynamics and provided a key framework for explaining gravitational phenomena. The observational advances, such as the direct Hubble parameter measurements \cite{Magana2018}, cosmic microwave background radiation (CMBR) \cite{Aghanim2020}, baryonic acoustic oscillations (BAO) \cite{Alam2017}, and supernovae type Ia (SNe Ia) \cite{Perlmutter1999,Riess1998} have shown that the universe is expanding, and doing so at an accelerated rate. 
\par According to the conventional $\Lambda$-cold dark matter ($\Lambda$CDM) model, this acceleration is caused by the cosmological constant ($\Lambda$) term in the field equations. This term yields an exotic fluid with negative pressure, which accounts for around $69\%$ of the universe's energy content. The $\Lambda$CDM model, while successful empirically, faces significant theoretical challenges. Two notable issues are the cosmological constant problem, which refers to the discrepancies between the observed and theoretically predicted values of the cosmological constant ($\Lambda$), and the coincidence problem, which questions why the current densities of dark energy and dark matter are so similar \cite{weinberg1989,padmanabhan2003,padilla2015,perivolaropoulos2008}. These unresolved issues have prompted the search for new theoretical frameworks capable of explaining cosmic acceleration without relying on a finely controlled cosmological constant.
\par Modifying the geometrical section of the Einstein field equations (EFE) is an alternate way to account for the missing matter-energy content in the visible cosmos.  The modified gravity theories \cite{Nojiri2017,Capozziello2011,Faraoni2011,Nojiri2011,Olmo2011,Capozziello2002,Carroll2004,Nojiri2003,Nojiri2007,Nojiri2008,Nojiri2005,Nojiri2005a,Nojiri2006a,Uddin2009,Boehmer2009,Alimohammadi2009,Cognola2007,Li2007} may explain the evolution of the cosmos. The \(f(R)\) theory \cite{Cembranos2009,Nojiri2004,Dunsby2010,Faraoni2008,Sotiriou2010,DeFelice2010} is one of the most fundamental extensions of GTR may be expressed by substituting the Ricci scalar \(R \) in the Einstein-Hilbert action with a generic function \(f(R) \).\\
A key motivation for using extended gravity theories to study cosmic acceleration is their ability to include quantum corrections through higher-order curvature terms. Various curvature invariants, such as \( R_{ij}R^{ij} \) and \( R_{ijkl}R^{ijkl} \), have been explored, but somelike the Kretschmann scalar, can produce unphysical solutions with growing anisotropy. Whereas\cite{Gorbunov2014,Myrzakulov2015,Bamba2014,Sebastiani2014}. The Gauss-Bonnet invariant $(\mathcal{G})$ combines the Ricci scalar, the Ricci tensor, and the Riemann tensor; it easily appears in quantum field theory when spacetime is curved.  The \(f(R,\mathcal{G}) \) framework, which includes both \(R \) and \(\mathcal{G}\), provides a comprehensive and unified extension of GTR by encompassing all significant curvature contributions. Based on the several studies, it has been observed that this modified gravity may describe a wide class of cosmological solutions \cite{Cognola2006,Elizalde2010,Myrzakulov2011,DeFelice2012,DeFelice2009a,DeFelice2011,DeFelice2010a,DeFelice2010b,DeFelice2010c}. These models may also imitate the $\Lambda$CDM model and other significant cosmological solutions of the early era \cite{Elizalde2010,Myrzakulov2011}.
\par  The $f(R,\mathcal{G})$ functional forms such as $f(R,\mathcal{G}) \equiv \alpha R^m + \beta \mathcal{G}^n $ and $f(R,\mathcal{G}) \equiv f_0 R^\delta  \mathcal{G}^\mu $ are allowed by the Noether symmetry method \cite{Bajardi2023}.  The dynamical system and observational analysis of a model having that $f(R,\mathcal{G}) \equiv f_0 R^\delta  \mathcal{G}^\mu $ form have been well explored \cite{DaCosta2018,bmishra,asingh2024,Bajardi2023}. We probe the dynamical evolution in $f(R,\mathcal{G}) \equiv \alpha R^m + \beta \mathcal{G}^n $ a model whose form is admitted in Noether symmetry analysis.
A further step in cosmological modelling is to explore potential interactions between dark matter and energy.  These interactions can overcome the coincidence problem and create unique signatures in the expansion of the universe's history \cite{Bolotin2015, Wang2016}. Specifically, the $f(R,\mathcal{G})$ gravity framework offers an ideal environment for investigating how interaction between modified gravity and dark sector interactions occurs by coupling a scalar field (as a dark energy candidate) to dark matter.  This technique has been prompted by recent reviews and studies that emphasize the importance of dark sector coupling in dealing with cosmological tensions as well as giving richer phenomena \cite{Bruck2015,Boehmer2015, Gleyzes2015,DAmico2016,Pan2020,Chatzidakis2022,asgrg2024,Gavela2009}.\\
The $f(R, \mathcal{G})$ gravity offers a comprehensive class of universe evolution regimes due to the non-linear contribution of $R$ and $\mathcal{G}$ appearing in the functional form of $f(R, \mathcal{G})$. An important motivation to analyze a particular functional form of $f(R, \mathcal{G})$ is its potential to describe the dynamical regimes of the universe. The application of dynamical system formulation allows us to evade the complexities of non-linearity to a good extent. Thus, the stability of asymptotic regimes in a flat FLRW universe with the interaction between dark matter and a scalar field may provide a more generalized as well as physically grounded framework to observe the dynamical cosmic evolution.
\par The $f(R,\mathcal{G})$ models suggest that the matter dominated phase may not be properly explained in this gravity for some class of $f(R,\mathcal{G})$ functions \cite{DaCosta2018,bmishra,asingh2024,Bajardi2023}. Using the additional degree of freedom through the scalar field $\phi$ and potential $V(\phi)$, we aim to describe the cosmological evolution in the $f(R,\mathcal{G})$ model. For a slowly evolving scalar field along with the potential, one may realize the cosmic acceleration responsible for dark energy or inflation. An oscillating scalar field around the minimum of potential may also be a candidate for the dark matter \cite{Tsujikawa2023}. In this sense the considered framework may be a remedy to explain the matter-dominated era in the $f(R,\mathcal{G})$ framework. The slowly rolling interacting scalar field may provide a dynamical mechanism to explain the matter-dominated era along with the effect of small effective cosmological constant. We probe whether the cosmological era composed of radiation, matter and late-time acceleration is realized in this framework or not. The possibility for these explanations is high due to the fact that the $f(R,\mathcal{G})$ model, when coupled with interacting scalar field and matter, may provide rich phenomenology as compared to the $f(R,\mathcal{G})$ theory illustrated in \cite{DaCosta2018,bmishra,asingh2024,Bajardi2023}.    
\par The present study is sequentially structured into separate sections as follows: Section \ref{sec2} is focused on presenting the key equations of $f(R, \mathcal{G})$ gravity framework, specifically addressing the action and field equations. In Section \ref{sec3}, we design a dynamical system, defining the relevant variables and write the equations describing their evolution. In the subsequent section \ref{sec4}, we analyze fixed points and their stability in the context of a specific $f(R,\mathcal{G}) =  \alpha R^m + \beta \mathcal{G}^n $ model. In Section \ref{sec5}, we employ statefinder diagnostic technique to analyze the characteristics associated with critical points. In section \ref{sec6}, we constrain the model parameters using observational Hubble data.  We provide an overview of results from our investigation in Section \ref{section6}.

\section{The basic equations of $f(R, \mathcal{G})$ gravity framework with interacting scalar field}
\label{sec2}
The interacting scalar field with matter introduces a non-trivial dynamics based on the energy transfer between matter and dark energy. Here, we study the consequences of interacting scalar field dynamics in the framework of $f(R, \mathcal{G})$ gravity. The action of $f(R, \mathcal{G})$ gravity with the canonical scalar field may be given by
\begin{equation} \label{eq1}
	S = \dfrac{1}{2\kappa^2}\int dx^4\sqrt{-g}[f(R,\mathcal{G}) + \mathcal{L}_m + \mathcal{L}_{\phi}].
\end{equation} 
It is worthwhile to mention that the above action (\ref{eq1}) would reduce to the $f(R,\mathcal{G})$ action \cite{DaCosta2018} for $\mathcal{L}_\phi=0$. Herein, $ f(R,\mathcal{G}) $  represents a function involving both the Ricci scalar ($R$), and the Gauss-Bonnet invariant $(\mathcal{G}) $. The standard matter Lagrangian density is denoted by $\mathcal{L}_m $. The Lagrangian of a canonical scalar field $\phi$ is denoted by $\mathcal{L}_{\phi}\equiv \frac{1}{2}(\triangledown \phi)^2-V(\phi)$ where $(\triangledown \phi)^2=g^{ij}\partial_i\partial_j\phi$. The scalar field \( \phi \) has been widely considered as a potential candidate for dark energy, capable of driving the late-time accelerated expansion of the universe. The energy density $\rho_\phi$ and pressure $p_\phi$, which will be defined explicitly later, enable diverse cosmic dynamics, including acceleration and deceleration phases \cite{Carroll1998,Nojiri2007a,Carvalho2006,B2023}.  \\
This study presents the scalar field as an independent component, rather than as a consequence of a scalar-tensor reformulation of the $f(R,\mathcal G)$ action. This method is frequently employed to examine the interaction between modified gravity and alternative dark energy sources. (see, e.g., \cite{Nojiri2006,Bamba2010}).  follow the natural units by setting \( c = k_B = \hbar = 1 \) and define \( \kappa^2 = 8\pi G\), with \( G \) being Newton’s gravitational constant. The explicit form of the Gauss-Bonnet term \( \mathcal{G} \) is 
\begin{equation}
	\mathcal{G} \equiv R^2-4R_{ij}R^{ij}+R_{ijkl}R^{ijkl}.
 \label{eqrev1}
\end{equation}
The Einstein's field equations corresponding to the action (\ref{eq1}) may be written by varying it with respect to the metric tensor field $ g_{ij} $ as
\begin{align}\label{s1}
	R_{ij} - \dfrac{1}{2} g_{ij}R = \kappa^2 T^{(m)}_{ij} + T^{(GB)}_{ij} + T^{(\phi)}_{ij},
\end{align}
It is worthwhile to mention that the above field equation (\ref{s1}) would reduce to the $f(R,\mathcal{G})$ field equations \cite{DaCosta2018} for $T^{(\phi)}_{ij}=0$. Here $T^{(m)}_{ij}$, $T^{(GB)}_{ij}$ and $T^{(\phi)}_{ij}$ are the stress-energy tensor for the matter, Gauss-Bonnet term and scalar field component, respectively. The stress-energy tensor for the standard matter is given by \cite{DaCosta2018}
\begin{equation}
	T^{(m)}_{ij} = -\dfrac{2}{\sqrt{-g}} \dfrac{\delta(\sqrt{-g}\mathcal{L}_m)}{\delta g_{ij}}. 
 \label{eqrev2}
\end{equation} 
and the energy-momentum term corresponding to $f(R,G)$ gravity as
\begin{align}
	T^{(GB)}_{ij} & = \nabla_i \nabla_j f_R - g_{ij} \square f_R+ 2R \nabla_i \nabla_j f_\mathcal{G} -2g_{ij}R \square f_\mathcal{G}-4R^\lambda_i \nabla_\lambda \nabla_j f_\mathcal{G}- 4R^\lambda_j \nabla_\lambda \nabla_i f_\mathcal{G} \nonumber \\
	& \quad  +4R_{ij}\square f_\mathcal{G}+4g_{ij} R^{kl}\nabla_k \nabla_l  f_\mathcal{G}+4R_{ijkl}  \nabla^k \nabla^l f_\mathcal{G} - \dfrac{1}{2} g_{ij}(Rf_R + \mathcal{G}  f_\mathcal{G} - f) \nonumber  \\
	& \quad +(1-f_R)\Big(R_{ij} -\dfrac{1}{2} g_{ij}R \Big).  
  \label{eqrev3}
\end{align}
where $ \square $ denotes the d'Alembert operator in the context of curved spacetime, and we use \begin{align}
	f_R \equiv \dfrac{\partial f(R, \mathcal{G})}{\partial R} \qquad {\rm and} \qquad f_\mathcal{G} \equiv \dfrac{\partial f(R, \mathcal{G})}{\partial \mathcal{G}}
  \label{eqrev4}
\end{align}
for the partial derivatives with respect to $ R $ and $ \mathcal{G} $. The present analysis involves the FLRW metric applied to a spatially flat universe ($k=0$), featuring a scale factor $a(t)$ that evolves over time. The metric is given by
\begin{equation} \label{eqn7}
	ds^2 = -dt^2 + a^2(t) [dx^2+dy^2+dz^2]
\end{equation}
In this case, the Ricci scalar and the Gauss-Bonnet invariant in the flat FLRW context are given by
\begin{align}
	R & = 6(2H^2 + \dot{H}), \label{eq8} \\ 
	\mathcal{G} &   =  24 H^2(H^2 + \dot{H}),\label{eq9}
\end{align}
The Hubble parameter, expressed as $H = \dot{a}/a$, where an overhead dot denotes the derivative with respect to cosmic time. Moreover, the field equations from equation (\ref{s1}) for the FLRW metric (\ref{eqn7}) would take the form
\begin{align} 
	3f_R H^2 
	&= \kappa^2(\rho_m + \rho_\phi) +\dfrac{1}{2}(f_R R - f -6H \dot{f_R} +\mathcal{G}f_\mathcal{G} - 24H^3\dot{f_\mathcal{G}}), \label{eq10} \\
	2f_R \dot{H}  &=  -\kappa^2( p_m + p_\phi + \rho_m + \rho_\phi) + H \dot{f_R} - \ddot{f_R} + 4H^3 \dot{f_\mathcal{G}} - 8H\dot{H} f_\mathcal{G} -4H^2 \ddot{f_\mathcal{G}} \label{eq11}
\end{align}
The Eqs. (\ref{eq10}) and (\ref{eq11}) can be rewritten by applying new definition as
\begin{align}
	3f_R H^2  &= \kappa^2(\rho_m + \rho_\phi + \rho_\Lambda) \label{eqa} \\ 
	2f_R \dot{H}  &=  -\kappa^2( \rho_m + p_m + p_\phi +  \rho_\phi +\rho_\Lambda + p_\Lambda). \label{eqb} 
\end{align}
The energy density ($ \rho_\phi $) and pressure ($ p_\phi $) for the canonical scalar field may be given by
\begin{align}
	\rho_\phi &= \dfrac{1}{2} \dot{\phi^2} + V(\phi) \label{eq12}\\
	p_\phi &= \dfrac{1}{2} \dot{\phi^2} - V(\phi).\label{eq13}
\end{align}
In order to understand the interacting scalar field dynamics in the $f(R,\mathcal{G})$ model, we proceed with the canonical scalar field with an exponential potential. This type of scalar field is frequently used to represent scalar fields in the universe \cite{Carroll1998,Nojiri2007a,Carvalho2006}. This form of potential, often linked to string theory, can't generate early inflation but influences how scalar field energy density evolves alongside matter and/or radiation \cite{Copeland1998}.  It may play a key role in dark energy models during the structure formation era, provided they respect nucleosynthesis bounds \cite{Copeland1998}.  Moreover, the exponential potential in cosmology may lead to power-law inflation in early universe, where the Hubble parameter may remain constant \cite{Rebesh2019,Halliwell1987,Neupane2004}. The exponentially accelerated expansion driven by the scalar field $\phi$ with sufficient flat potential $V(\phi)$ is the simplest and successful candidate of inflation \cite{Linde1982,Kamali2020,Guth1981}. The exponential potentials may also provide a viable cosmological mechanism for the explanation of accelerated expansion in the early universe as well as the late-universe \cite{Copeland1998,Heard2002,Arapoglu2019}. Motivated by these facts, we proceed with the following form of exponential potential as 
\begin{equation}
    V(\phi) = V_0  e^{-\lambda \phi}
     \label{eqrev9}
\end{equation}
where dimensionless parameters $\lambda, V_0 >0$. Even from the mathematical point of view, this form of potential often leads to the autonomous system during dynamical system formulation as well. Using Equations (\ref{eq12}) and (\ref{eq13}), the equation given above may be expressed as 
\begin{align}
	3f_R H^2  &=  \kappa^2\big(\rho_m + \dfrac{1}{2}\dot{\phi^2}+ V(\phi) + \rho_\Lambda \big) \label{eq1a} \\
	2f_R \dot{H}  &=  -\kappa^2( \rho_m + \dot{\phi}^2 +\rho_\Lambda + p_\Lambda) \label{eq1b}
\end{align}
From equations (\ref{eq10}), (\ref{eq11}), (\ref{eq1a}) and (\ref{eq1b}), we have
\begin{align}
	\kappa^2\rho_\Lambda &= \dfrac{1}{2}(f_R R - f -6H \dot{f_R} +\mathcal{G}f_\mathcal{G} - 24H^3\dot{f_\mathcal{G}}) \label{eq2a}\\
	-\kappa^2(\rho_\Lambda + p_\Lambda) &= H \dot{f_R} - \ddot{f_R} + 4H^3 \dot{f_\mathcal{G}} - 8H\dot{H}\dot{ f_\mathcal{G}} -4H^2 \ddot{f_\mathcal{G}} \label{eq2b}
\end{align}
In this way, the energy density $ \rho_\Lambda $ and pressure $ p_\Lambda $ of the geometrical dark energy may follow the conservation equation as
\begin{align}\label{eq21}
	\dot{\rho_\Lambda} = -3H(\rho_\Lambda + p_\Lambda).
\end{align}
where the equation of state parameter $ \omega_\Lambda = \dfrac{p_\Lambda}{\rho_\Lambda} $ is defined by
\begin{equation}
	\omega_\Lambda = \frac{p_\Lambda}{\rho_\Lambda} = -1-\frac{H \dot{f_R} - \ddot{f_R} + 4H^3 \dot{f_\mathcal{G}} - 8H\dot{H} \dot{f_\mathcal{G}} -4H^2 \ddot{f_\mathcal{G}}}{\dfrac{1}{2}(f_R R - f -6H \dot{f_R} +\mathcal{G}f_\mathcal{G} - 24H^3\dot{f_\mathcal{G}})}
  \label{eqrev10}
\end{equation}
We assume the interaction between the matter energy $\rho_m$ and scalar field $\rho_\phi$ \cite{COPELAND}. This is done by introducing the parameter $ Q $ which plays an important role in controlling the energy exchange rate in the dark sector. The energy moves from dark matter to dark energy in situations where $ Q > 0$. In contrast, the energy moves from dark energy to dark matter in situations where $ Q < 0$. The interaction between a quintessence scalar field $ \phi $ and dark matter, including its energy density $ \rho_m $, can be comprehensively depicted through balance equations
\begin{align}
	\dot{\rho_\phi}  &=  -3H(1+ \gamma_\phi)\rho_\phi + Q \label{eq3a} \\
	\dot{\rho_m}  &=  -3H\rho_m - Q. \label{eq3b}
\end{align}
Using equation (\ref{eq12}) and (\ref{eq13}) in equation (\ref{eq3a}), we may write the propagation equation of scalar field $ \phi $ with the dark sector coupling parameter $ Q $ as \cite{raushan2019dynamical,COPELAND}
\begin{align}
	\ddot{\phi} + 3H\dot{\phi} + \dfrac{dV}{d\phi} = \dfrac{Q}{\dot{\phi}}
\end{align}
We define the effective equation of state (EoS)parameter as
\begin{align}
	\omega_{eff} = -1 - \dfrac{2\dot{H}}{3H^2}.
\end{align}
In the effective scenario, we take the matter fluid (satisfying $p_m=0$) to be interacting with the scalar field satisfying $p_\phi=\gamma_\phi\rho_\phi$ in the $f(R,G)$ gravity framework. This set-up yields us the degree of freedom and helps us to reduce the non-linearity associated with the traditional $f(R,G)$ gravity field equations. It is worthwhile to mention that the $f(R,G)$ models may not always be written in the form of an autonomous system with non-linear exponents of the Gauss-Bonnet invariant \cite{asingh2024}. In the next section, we formulate the autonomous system in the model with the precise functional form $f(R,\mathcal{G})$ in the $f(R,\mathcal{G})$ gravity.
\section{Dynamical systems analysis of the Decoupled Power-Law $ f(R,G) $ model}
\label{sec3}
In this section, we use the dynamical system technique to study the universe evolution in model. In the dynamical system formulation of the cosmological models, the cosmological equations of the model are converted into an autonomous system. In the autonomous system, the independent variable does not appear explicitly \cite{asingh2023}. For the dynamical system given by $\dot{x}=f(x)$, where $x=(x_1,x_2,x_3,...,x_n)$, the critical points may be calculated by solving $\dot{x}=0$. The Jacobian matrix evaluated at critical points will possess eigenvalues. The eigenvalues are used to conclude on the stability of critical points of the cosmological dynamical system. If all the eigenvalues are of positive sign, then the critical point may be termed as an unstable point. This kind of point is also termed as a source or repeller of the system. In the phase space, the trajectories seem to diverge from the unstable point. If all the eigenvalues are of negative sign, then the critical point may be termed as the stable point. This kind of point is also known as the sink or attractor of the system. In the phase space, the trajectories seem to converge at the stable point. The points having eigenvalues of mixed signs are known as the saddle points. In the case of saddle points, the trajectories in the phase space would be converging along directions of negative signs and diverging along the directions having positive signs of eigenvalues \cite{Ellis1997,Coley2003}. \\
The dynamical system of the cosmological model may exhibit the cosmological phases of the universe's evolution corresponding to critical points. The attractor of a cosmological dynamical system reveals the accelerating universe expansion of the late-time era. The saddle points are suitable to portray the intermediate era of the universe's expansion, such as the matter-dominated era. The unstable critical points are suitable to portray the beginning era of the universe in a cosmological dynamical system \cite{Bahamonde2018}. We use these criteria to study the $f(R,\mathcal{G})$ model using dynamical systems. 
In order to study the $f(R,\mathcal{G})$ model using the dynamical system analysis, we define the  dimensionless variables as
\begin{align}\label{eq25}
	&	x_1 = \dfrac{\kappa^2 \rho_m}{3 f_R H^2}, \quad x_2 =\dfrac{\kappa^2 \dot{\phi^2}}{6 f_R H^2}, \quad x_3 =  \dfrac{\kappa^2 V(\phi)}{3 f_R H^2}, \quad x_4 = \dfrac{R}{6 H^2}, \quad x_5 = \dfrac{f}{6 f_R H^2},\nonumber   \\ &   x_6 = \dfrac{\dot{f_R}}{f_R H}, \quad x_7 = \dfrac{\mathcal{G} f_\mathcal{G}}{6 f_R H^2}, \quad x_8 = \dfrac{4H\dot{f_\mathcal{G}}}{f_R}.&
\end{align}
Using Eqs. (\ref{eq25}), (\ref{eq10}) and (\ref{eq12}), one may have
\begin{align}
	1 = x_1 + x_2 + x_3 + x_4 - x_5 - x_6 + x_7 - x_8 \label{eq26}
\end{align}
along with the density parameters $x_1 = \Omega_m$, $x_2 + x_3 = \Omega_\phi$, $x_4 - x_5 - x_6 + x_7 - x_8 = \Omega_\Lambda$. We use a dimensionless time variable $ N = {\rm ln} \,a(t) $ in the present expanding model. The formulation of the following dynamical system is achieved by computing the derivative of these variables with respect to $ N $. The dynamical system of the present framework may be written as
\begin{align}
		\frac{dx_1}{dN} &= -6 x_1 - x_1 x_6 -2x_1 \frac{\dot{H}}{H^2} \label{eq29a} \\
		\frac{dx_2}{dN} &= 3 \xi x_1 - 6x_2 + \lambda \frac{\dot{\phi}}{H}x_3 - x_2x_6 - 2x_2 \frac{\dot{H}}{H^2} \label{eq29b}\\ 
		\frac{dx_3}{dN} &=   -\lambda \frac{\dot{\phi}}{H}x_3 - x_3 x_6 -2x_3 \frac{\dot{H}}{H^2} \label{eq29c}\\  
         \frac{dx_4}{dN}  &=  \frac{\dot{R}}{6H^3}-2x_4 \frac{\dot{H}}{H^2} \label{eq29d}\\
		\frac{dx_5}{dN} &= \frac{\dot{f}}{6f_RH^3} - x_5 x_6 -2 x_5 \frac{\dot{H}}{H^2} \label{eq29e}\\
		\frac{dx_6}{dN} &= \frac{\ddot{f_R}}{f_R H^2} - x^2_6 - x_6 \frac{\dot{H}}{H^2} \label{eq29f}\\
		\frac{dx_7}{dN} &= \frac{\dot{\mathcal{G}}}{\mathcal{G}H} x_7 + \frac{\mathcal{G}}{24 H^4} x_8 - x_6 x_7 - 2 x_7 \frac{\dot{H}}{H^2}  \label{eq29g}\\
		\frac{dx_8}{dN} &= x_8 \frac{\dot{H}}{H^2} + \frac{4 \ddot{f_\mathcal{G}}}{f_R} - x_6 x_8   \label{eq29h}  
\end{align}
For the system to be closed, it is essential that all terms on the right-hand side of the mentioned equations are represented in terms of the variables indicated in equation (\ref{eq25}). Here, we consider $ Q = 3H \xi \rho_m $ \cite{Wang2016}. This interaction term being mathematically simple provides rich phenomenology in the cosmic dynamics of the model. The interaction, depending on matter energy density, leads to phenomenology based on clustering components in the universe. The matter that dominates during the early era may have strong effects during the deceleration era in the model based on $Q=3\xi H\rho_m$. The interaction parameter $\xi$ governs the direction of energy transfer during expanding evolution.  From equations (\ref{eq8}), (\ref{eq9}), and (\ref{eq11}), we determine the necessary expressions as $\frac{\dot{H}}{H^2} =  x_4 - 2$, $\frac{\dot{R}}{6H^3}   =   \frac{x_4 x_6}{b}$ and $\frac{\mathcal{G}}{24 H^4} =  x_4 - 1$ alon with
\begin{align}
	\frac{\dot{f}}{6f_RH^3}  & =  \frac{x_4 x_6}{b} + \dfrac{x_7}{x_4-1} \Big[\frac{x_4 x_6}{b}+2(x_4-2)^2\Big]	\label{eq29j}\\
	\frac{\dot{\mathcal{G}}}{\mathcal{G}H}   &  =  \frac{1}{(x_4 - 1)}  \Big[\frac{x_4 x_6}{b} + 2(x_4 -2)^2 \Big]  \label{eq29l}\\ 
	\frac{4\ddot{f_\mathcal{G}}}{f_R}  & =  -6 x_2 - 3x_1 + x_6 + x_8 (5 - 2 x_4) - 2(x_4 - 2) - \frac{\ddot{f_R}}{f_R H^2}\label{eq29n} .
\end{align}
Using equations (\ref{eq29j}-\ref{eq29n}), the system (\ref{eq29a}-\ref{eq29h}) may be transformed as
\begin{align}
		\frac{dx_1}{dN} &= -3(1+\xi)x_1 - x_1 x_6 -2x_1 (x_4 - 2) \label{eq28a}\\ 
        \frac{dx_2}{dN} &= 3 \xi x_1 - 6x_2 + \lambda \frac{\dot{\phi}}{H}x_3 - x_2x_6 - 2x_2 (x_4 - 2) \label{eq28b}\\	
        \frac{dx_3}{dN} &=   -\lambda \frac{\dot{\phi}}{H}x_3 - x_3 x_6 -2x_3 (x_4 - 2)\label{eq28c} \\
        \frac{dx_4}{dN}  &=  \frac{x_4 x_6}{b}-2x_4 (x_4 - 2)	\label{eq28d}\\
        \frac{dx_5}{dN} &= \frac{x_4 x_6}{b} + \dfrac{x_7}{x_4-1}\Big[\frac{x_4 x_6}{b}+2(x_4-2)^2\Big] - x_5 x_6 - 2 x_5(x_4-2) \label{eq28e}\\
		\frac{dx_6}{dN} &= \Gamma - x^2_6 - x_6(x_4 -2) \label{eq28f}\\
		\frac{dx_7}{dN} &= \frac{x_7}{x_4 -1} \Big[\frac{x_4 x_6}{b} + 2(x_4 -2)^2 \Big] + (x_4 -1) x_8 - x_7 x_6 -2 x_7(x_4-2) \label{eq28g} \\
		\frac{dx_8}{dN} &=- x_8(x_4 + x_6 -3) -3x_1 -6x_2 + x_6 -2(x_4 -2)- \Gamma \label{eq28h}
\end{align}
where $b = \frac{d\ln f_R}{d\ln R} = \frac{Rf_{RR}}{f_R}$. These equations would govern the cosmological evolution within a generalized $ f(R, \mathcal{G}) $ gravity theory, with the specific nature of the theory describe by $ \Gamma = \dfrac{\ddot{f_R}}{f_R H^2} $. Furthermore, we also have $\omega_{eff}  = -\dfrac{1}{3}(2x_4 -1)$ and $
 \omega_\Lambda = -1-\dfrac{3x_1 +6x_2+2x_4-4}{3(1-x_1 - x_2 - x_3)}$. Additionally, the deceleration parameter is obtained as $q= \dfrac{d}{dt}\Big(\dfrac{1}{H} \Big)-1 = -1- \dfrac{\dot{H}}{H^2} = 1-x_4$. \\
As a general rule, the system is considered open until the expression for $ \Gamma $ is presented in terms of the dynamical variables (\ref{eq25}). \\  
In other words, we proceed with the choice of  $f(R,\mathcal{G}) \equiv \alpha R^m + \beta \mathcal{G}^n $  characterized by parameters $\alpha, \beta, m, $ and  $ n $ gravity and we choose the coupling parameter $Q=3H \xi \rho_m$ in our calculations. This model stands as a generalization of different gravity theories. Specifically, setting  $ \alpha \neq 0 , \beta = 0,$ and $ m=1 $ yields Einstein's gravity,  while  $ \alpha \neq 0, \beta = 0,$ and $ m =2 $ corresponds to  $ R^2 $ gravity, and  $ \alpha \neq 0, \beta = 0$ results in $ f(R).$ Furthermore, in the case of  $ \alpha = 0, \beta \neq 0, $ the model transforms into $ f(\mathcal{G})$ gravity. Moreover, we consider $ \phi = \ln H^{m_1} $  where $H$ is the Hubble parameter and  $ m_1 $ is an arbitrary positive real number.  Consequently, the relation $ \dfrac{\lambda\dot{\phi}}{H} = \delta (x_4 - 2) $ may be obtained, where $ \delta = \lambda. m_1 $ is a constant. Specifically, we may have
\begin{eqnarray} 
b   =  (m -1), \quad x_5   =  \dfrac{x_4}{m} + \dfrac{x_7}{n}, \quad x_8   =  \dfrac{(n-1)x_7}{(x_4-1)^2} \Big[\dfrac{x_4 x_6}{b} + 2(x_4 -2)^2 \Big] \label{eqCC}\\
	x_6 = \dfrac{-1+ x_1 + x_2 + x_3 + x_4 \Big(\frac{m-1}{m}\Big) +\dfrac{x_7 (n-1)}{(x_4-1)^2} \Big[\frac{(x_4-1)^2}{n} - 2(x_4-2)^2\Big]}{1 + \frac{(n-1)x_4 x_7 }{b(x_4-1)^2}} \label{eqDD}.
\end{eqnarray}
By using the above equations (\ref{eqCC}-\ref{eqDD}), we may thus rule out $ x_5, x_6 $, and $ x_8 $, since these variables depend on other variables. From Eqs. (\ref{eq28a}-\ref{eq28h}), The autonomous system of the model would take the form
\begin{align}
		\frac{dx_1}{dN} &=  -3(1+\xi)x_1 - x_1 x_6 -2x_1 (x_4 - 2) \label{eq43a}\\ 
		\frac{dx_2}{dN} &=  3 \xi x_1 - 6x_2 + \delta (x_4 -2)x_3 - x_2x_6 - 2x_2 (x_4 - 2) \label{eq43b} \\
		\frac{dx_3}{dN} &=   -\delta (x_4 -2)x_3 - x_3 x_6 -2x_3 (x_4 - 2) \label{eq43c} \\
		\frac{dx_4}{dN} &=   \frac{x_4 x_6}{m-1}-2x_4 (x_4 - 2)	\label{eq43d} \\
		\frac{dx_7}{dN} &= \frac{x_7}{x_4 -1} \Big[\frac{x_4 x_6}{b} + 2(x_4 -2)^2 \Big] + (x_4 -1) x_8 - x_7x_6 -2 x_7(x_4-2) \label{eq43e}
    \end{align}
We may obtain the critical points of the above system and determine their stability.
\section{The critical points and their cosmological implications}\label{sec4}
For getting the fixed points, we equate the autonomous system (\ref{eq43a}-\ref{eq43e}) to zero. The fixed points coordinates $ (x_1, x_2, x_3, x_4, x_7)  $ and their corresponding existence conditions are given in Table \ref{t1}. This table also incorporates the hyperbolicity conditions of the critical points. The summary of eigenvalues at the critical points are given in Table \ref{t2}.
\begin{table}[h!]
    \centering
    \resizebox{\textwidth}{!}{%
    \begin{tabular}{|c|c|c|c|}
      \hline
      Critical points & $(x_1, x_2, x_3, x_4, x_7)$ & Existence & Hyperbolicity \\
      \hline
      A & $(0, 0, 0, 0, 0)$ & Always &
      \begin{tabular}{c}
      $\xi = \tfrac{2}{3}$  Non-hyperbolic\\
      $m= \tfrac{5}{4}$ Non-hyperbolic\\
      $n = \tfrac{5}{8}$  Non-hyperbolic\\
      otherwise  Hyperbolic
      \end{tabular} \\ [6ex]
      B & $(0, -1, 0, 0, 0)$ & Always &
      \begin{tabular}{c}
      $\xi = 1$  Non-hyperbolic\\
      $m= \tfrac{3}{2}$ Non-hyperbolic\\
      $n = \tfrac{3}{4}$  Non-hyperbolic\\
      otherwise  Hyperbolic
      \end{tabular} \\ [6ex]
      C & $((\xi-1)(3\xi-2), \xi(2-3\xi), 0, 0, 0)$ & Always &
      \begin{tabular}{c}
      $\xi =\tfrac{2}{3}, 1$ and $-\tfrac{(2\delta+3)}{3} $  Non-hyperbolic\\
      $m= \tfrac{3}{4}(\xi+1)$ Non-hyperbolic\\
      $n = \tfrac{3}{8}(\xi+1)$  Non-hyperbolic\\
      otherwise  Hyperbolic
      \end{tabular} \\ [6ex]
      D & $(0, 0, 0, 0, \frac{8n^2-5n}{(8n-1)(n-1)})$ & $n \neq 1,\frac{1}{8}$ &
      \begin{tabular}{c}
      $m= 2n$ Non-hyperbolic\\
      $n = \tfrac{5}{8}, \tfrac{3}{4}, \tfrac{3}{8}(\xi+1)$ and $-\tfrac{\delta}{4}$  Non-hyperbolic\\
      otherwise  Hyperbolic
      \end{tabular} \\ [6ex]
      E & $\Big(0, \frac{-\delta(2\delta+5)}{3}, \frac{(2\delta+5)(\delta+3)}{3}, 0, 0\Big)$ & Always &
      \begin{tabular}{c}
      $\xi =-\tfrac{(2\delta+3)}{3} $  Non-hyperbolic\\
      $m= -\tfrac{\delta}{2}$ Non-hyperbolic\\
      $n = -\tfrac{\delta}{4}$  Non-hyperbolic\\
      otherwise  Hyperbolic
      \end{tabular} \\ [6ex]
      F & $\Big(0, 0, \frac{2-m}{m}, 2, 0 \Big)$ & $m \neq 0$ & Hyperbolic \\ [2ex]
      G & $\Big(x_{1G}, x_{2G}, 0, \frac{-(3\xi-4m+3)}{2m}, 0\Big)$ & $m \neq 0$ &
      \begin{tabular}{c}
      $\xi = 1$ and $-1$  Non-hyperbolic\\
      $m= 2n$ and $-\tfrac{\delta}{2}$ Non-hyperbolic\\
      otherwise  Hyperbolic
      \end{tabular} \\ [6ex]
      H & $\Big(0, \frac{-(7m^2-11m+3)}{m^2}, 0, \frac{2m-3}{m}, 0 \Big)$ & $m \neq 0$ &
      \begin{tabular}{c}
      $\xi = 1$  Non-hyperbolic\\
      $m= -\tfrac{\delta}{2}$ and $m=2n$ Non-hyperbolic\\
      otherwise  Hyperbolic
      \end{tabular} \\ [4ex]
      I & $\Big(0, 0, 0, 2, \frac{-(2n-mn)}{m-mn} \Big)$ & $m \neq 0, n \neq 1$ & Non-hyperbolic \\ [4ex]
      J & $\Big(0, 0, 0, \frac{4m^2-5m}{(2m-1)(m-1)}, 0\Big)$ & $m \neq 1, 1/2$ &
      \begin{tabular}{c}
      $m= 2,\tfrac{5}{4}$ and $-\tfrac{\delta}{2}$ Non-hyperbolic\\
      $n = \tfrac{m}{2}$  Non-hyperbolic\\
      otherwise  Hyperbolic
      \end{tabular} \\ [6ex]
      \hline
    \end{tabular}%
    }
    \caption{The critical points and their corresponding effective equation of state parameter, where, $x_{1G} = \dfrac{(\xi-1)(6m^2\xi+8m^2-9m\xi-13m+3\xi+3)}{2m^2}$ \& $x_{2G} = \dfrac{-\xi(6m^2\xi+8m^2-9m\xi-13m+3\xi+3)}{2m^2}$.}
    \label{t1}
\end{table}

\begin{table}[h!]
    \centering
    \begin{tabular}{|c|c|c|}
        \hline
 Critical points & Eigenvalues & Stability nature \\ 
    \hline
         A &  $\left[-1, 2-3\xi ,2\delta+5, 5-8n, \frac{4m-5}{m-1} \right]$ & Saddle \\ [2ex]
           B &  $\left[1,3-3\xi,2\delta+6,6-8n,\frac{4m-6}{m-1}\right]$ & Unstable or Saddle \\[2ex]
            C &  $\left[3\xi-2,3(\xi-1),(3\xi-8n+3),(2\delta+3\xi+3),\frac{4m-3\xi-3}{m-1}\right]$ & Stable or Unstable \\[2ex]
           D &  $\left[8n-5,8n-6,8n-3\xi-3,\frac{4m-8n}{m-1},2\delta+8n\right]$ &  Stable or Unstable \\[2ex]
           E &   $\left[-2\delta-5,-2\delta-6, -2\delta-3\xi-3,\frac{2\delta+4m}{m-1},-2\delta-8n\right]$ & Saddle or Stable \\[2ex]
           F &  $\lambda_F$ & Saddle\\[2ex]
           G &  $\left[3(\xi-1),\frac{3(\xi+1)(m-2n)}{m},\frac{3(\xi+1)(\delta+2m)}{2m},\lambda_{4G},\lambda_{5G}\right]$ & Saddle\\[2ex]
           H & $\left[\frac{3\delta+6m}{m},\frac{6m-12n}{m}, 3-3\xi,\lambda_{4H}, \lambda_{5H}\right]$ & Saddle\\[2ex]
            I & $\left[-6,0,-3(1+\xi),\lambda_{4I}, \lambda_{5I}\right]$  & Saddle\\[2ex]
           J & $\left[\frac{-(\delta+2m)(m-2)}{(2m-1)(m-1)},\frac{-2(m-2n)(m-2)}{(2m-1)(m-1)}, \frac{5-4m}{m-1} ,\lambda_{4J}, \lambda_{5J}\right]$ & Stable or Unstable \\[2ex]
        \hline  
    \end{tabular}
    \caption{The eigenvalues corresponding to the critical points, where $\lambda_F$ has very large values and $\lambda_{4G} = \frac{-3(1 + \xi - 2m\lambda)}{4m} + \frac{\sqrt{4m^3(3\xi + 8)^2 - 4m^2(165\xi + 54\xi^2 + 152) + 3m(87\xi + 139)(\xi + 1) - 81(\xi + 1)^2}}{4m(m - 1)^{1/2}}$,  $\lambda_{5G} = \frac{-3(1 + \xi - 2m\lambda)}{4m} 
- \frac{\sqrt{4m^3(3\xi + 8)^2 - 4m^2(165\xi + 54\xi^2 + 152) + 3m(87\xi + 139)(\xi + 1) - 81(\xi + 1)^2}}{4m(m - 1)^{1/2}}$ \\ $\lambda_{4H} =  -\frac{(121m^3-371m^2+339m-81)^{1/2} -3(m-1)^{3/2}}{2m(m-1)^{1/2}} $, $\lambda_{5H} =  \frac{(121m^3-371m^2+339m-81)^{1/2} +3(m-1)^{3/2}}{2m(m-1)^{1/2}} $, \\ $ \lambda_{4I} = \frac{-1}{2} (\frac{41m-100n+50mn-25n^2}{m-4n+2mn-m^2})^{1/2}- \frac{3}{2} $, $ \lambda_{5I} = \frac{1}{2} (\frac{41m-100n+50mn-25n^2}{m-4n+2mn-m^2})^{1/2}- \frac{3}{2} $,\\ $\lambda_{4J}= \frac{-(8m^2-13m+3)}{2m^2-3m+1} -3\xi$ $\lambda_{5J}= \frac{-(14m^2-22m+6)}{2m^2-3m+1}$.}
    \label{t2}
\end{table}

The system (\ref{eq43a}–\ref{eq43e}) contains ten critical points.  The stability and cosmological characteristics of these critical points are examined with respect to cosmological parameters, including the deceleration parameter, \( q \), the EoS \( \omega_{\text{eff}} \), the Hubble parameter, \( H \), and the scale factor, \( a \), as detailed in Table \ref{t3}. The cosmological implications of these points are as follows:
\par \textit{Point A}: This ever-existing point acts like a saddle point for all values of $ m,n $ and $ \xi $ with the exception of $ m \neq 1, 5/4,  n \neq 5/8  $ and $\xi \neq 2/3.$
This point corresponds to the decelerating expansion of the universe influenced by a radiation-like fluid. The universe will expand with $ a \propto t^{1/2}$. This point lies at the origin of $5D$ phase space. The cosmic dynamics at this point will be governed by the effective fluid behaving like radiation in the model.
\par \textit{Point B}: Point B will always exist in the cosmological dynamical system. The behavior of this point is like a repeller for $m <1 $ or $m > 3/2, n<3/4$ and $\xi<1$ and for $m \in(1,3/2), n>3/4$ and $\xi>1$ it behaves like a saddle point. This refers to the slowing expansion of the cosmos affected by a radiation-like fluid. The decelerating expansion phase corresponding to this point will have $f_R<0$. It means that the behavior of $f(R,\mathcal{G})$ will be decreasing with respect to the increasing Ricci scalar. This kind of dynamics may be possible during the radiation phase governed by the curvature terms based on $R$ and $\mathcal{G}$. The universe will expand according to the relation $ a \propto t^{1/2}$ leading to the Hubble parameter $H\propto \frac{1}{2t}$. 
\par \textit{Point C}: This point will be ever-existing. The point corresponds to the non-zero contribution from the matter density $(\rho_m)$ and the kinetic term $(\dot{\phi}^2)$. The effective scenario due to these contributions leads to the cosmic expansion governed by the effective radiation fluid. For cases $ \big(\xi <\frac{1}{3}\cap \frac{1}{4} (3 \xi +3)<m<1\cap \delta<\frac{1}{2} (-3 \xi -3)\cap n>\frac{1}{8} (3 \xi +3)\big) $ or $\big(\frac{1}{3}<\xi <\frac{2}{3}\cap 1<m<\frac{1}{4} (3 \xi +3)\cap \delta<\frac{1}{2} (-3 \xi -3)\cap n>\frac{1}{8} (3 \xi +3)\big)$, the point behaves like an attractor. In cases where $\xi >1$ and $\left(m<1\cap \delta>\frac{1}{2} (-3 \xi -3)\cap n<\frac{1}{8} (3 \xi +3)\right)\cup \left(m>\frac{1}{4} (3 \xi +3)\cap \delta>\frac{1}{2} (-3 \xi -3)\cap n<\frac{1}{8} (3 \xi +3)\right)$, the point will act as a repeller.  This is the point at which the expansion of the cosmos is slowing under the influence of a radiation-like effective fluid. Furthermore, in this scenario, the value of the scale factor is given by $ a \propto  t^{1/2}$ . 
\par \textit{Point D}: The point  D will always exist except $ n= 1/8, 1 $.  For the cases \begin{align*}
\left(n < \frac{1}{2} \cap 2n < m < 1 \cap \delta < -4n \cap \xi > \frac{1}{3} (8n - 3)\right) \, \text{or} \, \\
\left(\frac{1}{2} < n < \frac{5}{8} \cap 1 < m < 2n \cap \delta < -4n \cap \xi > \frac{1}{3} (8n - 3)\right),
\end{align*}
the point behaves like a stable point. In the same way, the cases where,\begin{align*}
\left(\xi \leq 1 \cap n > \frac{3}{4} \cap \delta > -4n \cap (m < 1 \cup m > 2n)\right) \, \text{or} \, \\
\left(\xi > 1 \cap n > \frac{1}{8} (3 \xi + 3) \cap \delta > -4n \cap (m < 1 \cup m > 2n)\right), \end{align*} the point will act like an unstable point. The point $D$ belongs to the plane governed by Gauss-Bonnet invariant $\mathcal{G}$. The curvature term based effective dynamic evolution leads to the universe expansion scenario based on the radiation-like effective fluid.  In the presence of a radiation-type fluid, the expansion rate of the universe slows, and its scale factor follows the relation $ \ a\propto t^{1/2}$.
\par \textit{Point E}:  This ever-existing point E will demonstrate stable behavior for
\begin{align*}
&\left(\xi \leq \frac{1}{3} \cap \delta > \frac{1}{2} (-3 \xi - 3) \cap -\frac{\delta}{2} < m < 1 \cap n > -\frac{\delta}{4}\right) \, \cup \, \\
&\left(\xi > \frac{1}{3} \cap \left(\left(\frac{1}{2} (-3 \xi - 3) < \delta < -2 \cap 1 < m < -\frac{\delta}{2} \cap n > -\frac{\delta}{4}\right) \, \cup \right.\right. \\
&\left.\left(\delta > -2 \cap -\frac{\delta}{2} < m < 1 \cap n > -\frac{\delta}{4}\right)\right)\Bigg)
\end{align*}
And, when conditions \begin{align*}
&\left(\xi < \frac{1}{3} \cap \left(\left(\delta \leq -2 \cap n < -\frac{\delta}{4} \cap \left(m < 1 \cup m > -\frac{\delta}{2}\right)\right) \, \cup \right.\right. \\
&\left.\left(-2 < \delta < \frac{1}{2} (-3 \xi - 3) \cap n < -\frac{\delta}{4} \cap \left(m < -\frac{\delta}{2} \cup m > 1\right)\right)\right)\Bigg) \, \cup \, \\
&\left(\xi \geq \frac{1}{3} \cap \delta < \frac{1}{2} (-3 \xi - 3) \cap n < -\frac{\delta}{4} \cap \left(m < 1 \cup m > -\frac{\delta}{2}\right)\right)
\end{align*}
are fulfilled, then this point will exhibit saddle behavior. This point has non-zero $x_2$ and $x_3$ terms and thus it belongs to the $x_2-x_3$ plane governed by the kinetic and potential terms of the scalar field. Although, the potential term used to be responsible for the dark energy dynamics in the non-interacting scenarios, in this model the contribution from curvature terms based on $R$ and $\mathcal{G}$ are so strong that the effective dynamics corresponding to this point is like the radiation-dominated universe. This point corresponds to the decelerating expansion of the universe influenced by a radiation-like fluid. The universe will expand with $ \ a\propto t^{1/2}$. It is an interesting aspect of this model.
\par \textit{Point F}: The critical point F will always be there in the model except for $m= 0$. At this point, the universe is mainly influenced by dark energy and indicates the de Sitter universe with exponential expansion, given by the scale factor $a \propto e^{H_0 t}$. Stability analysis is not achievable at this point since the eigenvalue expressions turn out to be too lengthy. The $\Lambda$CDM phase having $q=-1,\omega_{eff}=-1$, $H=$ constant will be characterized corresponding to this point in the model. The non-zero potential term ($x_3$) due to the interacting scenario, along with the Ricci scalar based $x_4$ term, is responsible for the de Sitter expansion scenario in the cosmological dynamical system.
\par \textit{Point G}: The critical point G is always there in the model except for $m = 0$. Since the stability conditions of the respective eigenvalues are very long, we chose not to include them. However, the analysis suggests that the point could be stable, unstable, or a saddle, depending on the different parameters of the model. At this point, we found that different values of $m$ and $\xi$ result in varying cosmological scenarios.A matter-dominated expansion is realised when the value of   $m = 1 + \xi$ and the choice $m = \frac{3(1+\xi)}{4}$ results in the universe exhibiting radiation-like behavior.  When $\xi = -1$, dark energy takes over. Similarly, for $m < \frac{3(1+\xi)}{2}$, the universe is decelerated, while for $m > \frac{3(1+\xi)}{2}$, it is accelerated. In this case, the relationship for the scale factor follows $a \propto  t^{\frac{2m}{3(\xi+1)}}$. These different cosmic phases will not have contributions from the potential term and Gauss-Bonnet term based dynamical variables $x_3$ and $x_7$ respectively. This point is appropriate to explain the matter-dominated era of the universe's evolution in this model.
\par \textit{Point H}: The critical point H will exist whenever $ m \neq 0 $. This point shows the saddle behavior for $ m > 1 $ and $ n< m/2$. The non-zero contribution from the kinetic term of the scalar field and Ricci scalar governs the cosmic dynamics at this point. In this case, the expansion of the universe is marked by the scale factor $ a \propto t^{m/3} $. As a result, if the value of $ m $ satisfies the range $m <3 $, the cosmos will experience a decelerating phase of expansion and for $ m>3 $, the universe will undergo an accelerating phase of expansion. In this scenario, the condition  $ m=2 $ corresponds to a cosmological phase dominated by matter. Radiation becomes the dominant component when the parameter $m = 3/2$.
\par \textit{Point I}: Under the constraint $ n\neq 1, m \neq 0 $, the critical point I continues to exist in the dynamical cosmological system. Conducting a stability analysis is currently impractical because the resulting eigenvalue expressions are too intricate and lengthy for direct interpretation. In this particular case, the expansion of the universe is identified by the scale factor $ a \propto e^{H_0 t}, $ where $ H_0$ is constant. Moreover, the value of the deceleration parameter represents the de Sitter universe that has exponential expansion. Consequently, the EoS parameter represents a specific situation acknowledged as the cosmological constant or the vacuum energy scenario. The cosmic expansion at this point is qualitatively similar to the point $F$. However, it is to be noted that the cosmic dynamics at this point is purely controlled by the curvature based $R$ and $\mathcal{G}$ terms belonging to the $f(R,\mathcal{G})$ gravity. In the cosmic dynamics corresponding to this point, one may not have the role of interacting scalar field dynamics. It is an important implication of this point.
\par \textit{Point J}: Under the constraint $ m\neq 1, 1/2 $ , the critical point J will exist in the dynamical cosmological system. It is noticed that point J exhibits an unstable nature for \begin{align*}
&\left(n \leq \frac{1}{2} \cap 1 < m < \frac{1}{14} \left(\sqrt{37} + 11 \right) \cap \delta > -2m \cap \xi < \frac{-8m^2 + 13m - 3}{6m^2 - 9m + 3}\right) \, \text{ or } \, \\
&\left(\frac{1}{2} < n < \frac{1}{28} \left(\sqrt{37} + 11 \right) \cap 2n < m < \frac{1}{14} \left(\sqrt{37} + 11 \right) \cap \delta > -2m \cap \xi < \frac{-8m^2 + 13m - 3}{6m^2 - 9m + 3}\right).
\end{align*}
 It will act like an attractor otherwise. The expansion of the universe is identified through the scale factor that is governed by the expression $ a \propto t ^{\frac{(m-1)(2m-1)}{(2-m)t}}$ .\\
In addition to this, the universe will undergo an accelerating expansion for $ (m<\frac{1}{2} \left(1-\sqrt{3}\right)\cup \frac{1}{2}<m<1)\cup \big(m>\frac{1}{2} \left(\sqrt{3}+1\right)\big)$ and a deceleration for $( \frac{1}{2} \left(1-\sqrt{3}\right)<m<\frac{1}{2}) \cup  \big(1<m<\frac{1}{2} \left(\sqrt{3}+1\right)\big)$. Also, if the value of $ m $ falls within the range $ (\frac{1}{2}<m<1)\cup m>2  $, the point will be governed by a quintessence kind of dark energy and, for that $ m<\frac{1}{2}\cup (1<m<2) $, it will have a phantom kind of dark energy. Additionally, the model predicts a matter-dominated phase for the parameter choice  $ m = \dfrac{7 \pm \sqrt{73}}{12} $  and radiation dominates the cosmic behavior for the specific values $m= 0, 5/4$. The value of $ \omega_{eff} = -1  $  at $m=-1$ In this case, the universe is dominated by the cosmological constant-like dark energy.\\
\begin{table}[htbp]
\centering
\small
\caption{The cosmological quantities at the fixed points of power law model.}
\resizebox{\textwidth}{!}{%
\begin{tabular}{lccccccc}
\toprule
Point & $q$ & $\omega_{\text{eff}}$ & $H$ & $a$ & $r$ & $s$ \\
\midrule
A  & 1 & $ \frac{1}{3} $ & $\frac{1}{2t}$ & $t^{1/2}$ & 3 & $\frac{4}{3}$ \\
B  & 1 & $ \frac{1}{3} $ & $\frac{1}{2t}$ & $t^{1/2}$ & 3 & $\frac{4}{3}$ \\
C  & 1 & $ \frac{1}{3} $ & $\frac{1}{2t}$ & $t^{1/2}$ & 3 & $\frac{4}{3}$ \\
D  & 1 & $ \frac{1}{3} $ & $\frac{1}{2t}$ & $t^{1/2}$ & 3 & $\frac{4}{3}$ \\
E  & 1 & $ \frac{1}{3} $ & $\frac{1}{2t}$ & $t^{1/2}$ & 3 & $\frac{4}{3}$ \\
F  & -1 & -1 & $H_0$ & $e^{H_0 t}$ & 1 & 0 \\
G  & $\frac{3\xi+3-2m}{2m}$ & $\frac{\xi+1-m}{m}$ & $\frac{2m}{3(\xi+1)t}$ & $t^{\frac{2m}{3(\xi +1)}}$ & 
$\frac{(-2m + 3\xi + 3)(-m+3\xi+3)}{2m^2} $ & 
$\frac{-1 + \frac{-2m + 3\xi + 3}{2m} + \frac{(-2m + 3\xi + 3)^2}{2m^2}}{-1.5 + \frac{3(-2m + 3\xi + 3)}{2m}}$ \\
H  & $\frac{3-m}{m}$ & $\frac{2-m}{m}$ & $\frac{m}{3t}$ & $t^{\frac{m}{3}}$ & $\frac{m^2 - 9m + 18}{m^2}$ & $\frac{2}{m}$ \\
I  & -1 & -1 & 1 & $e^{H_0 t}$ & 1 & 0 \\
J  & $\frac{-2m^2 + 2m + 1}{2m^2 - 3m + 1}$ & 
$-\frac{1}{3} \cdot \frac{6m^2 - 7m - 1}{2m^2 - 3m + 1}$ & 
$\frac{(m - 1)(2m - 1)}{(2 - m)t}$ & 
$t^{\frac{(m - 1)(2m - 1)}{2 - m}}$ & 
$\frac{4m^4 - 6m^3 - 6m^2 + 7m + 3}{4m^4 - 12m^3 + 13m^2 - 6m + 1}$ & 
$\frac{2(2 - m)}{3(2m^2 - 3m + 1)}$ \\
\bottomrule
\end{tabular}%
}
\label{t3}
\end{table}

\begin{figure}[h!]
\centering
		\includegraphics[height=8cm,width=8cm]{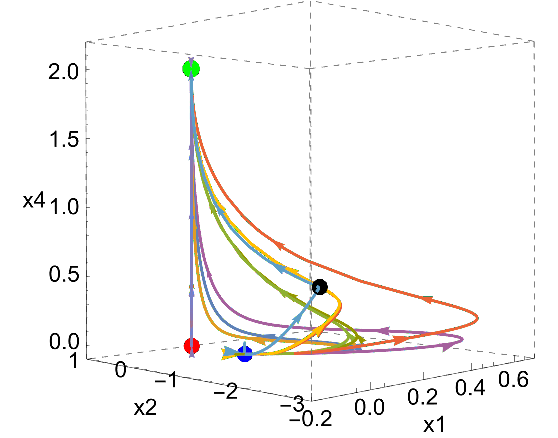}
		\caption{ 3D Phase portrait in $ x_1-x_2-x_4 $ plane for $ m =2, \xi =1, \delta \in \mathbb{R}^{+} $ and $n \in \mathbb{R} $  with coordinates of red point $(0,0,0)$, blue point $(0,-1,0)$, green point $(0,0,2)$, and black point $(0,-\frac{9}{4},\frac{1}{2})$. } 
\label{f2}
\end{figure}

\begin{figure}[h!]
\centering
		\includegraphics[height=8cm,width=8cm]{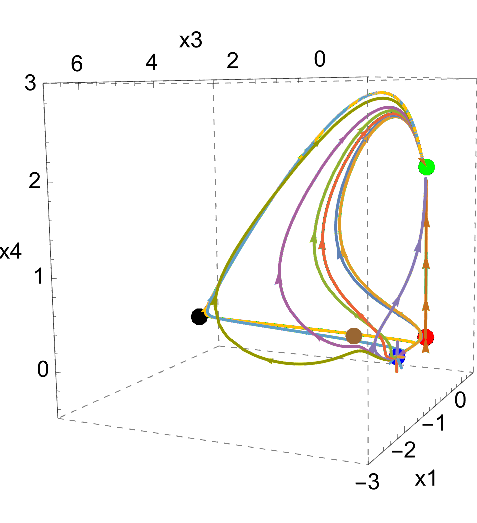}
		\caption{ 3D Phase portrait in $ x_1-x_3-x_4 $ plane for $ m =2, \xi = 1, \delta=1,$ and $n \in \mathbb{R} $ with coordinates of red point $(0,0,0)$, blue point $(-1,0,0)$, green point $(0,0,2)$, black point $(0,7,0)$, and brown point $(-\frac{9}{4},0,\frac{1}{2})$.}
\label{f3}
\end{figure}

\begin{figure}[h!]
\centering
		\includegraphics[height=9cm,width=9cm]{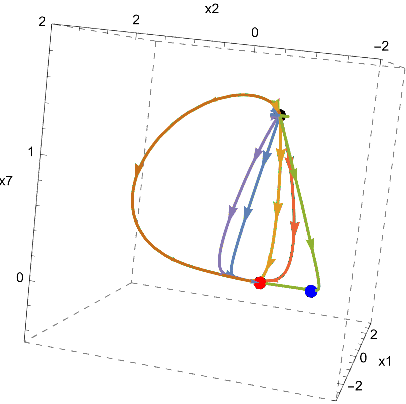}
		\caption{ 3D Phase portrait in $ x_1-x_2-x_7 $ plane for $ n =2, \xi =1, \delta \in \mathbb{R}^{+} $ and $m \in \mathbb{R}$ with coordinates of  red point $(0,0,0)$, blue point $(0,-1,0)$, and black point $(0,0,\frac{22}{15})$.} 
\label{f4}
\end{figure}

\begin{figure}[h!]
\centering
		\includegraphics[height=9cm,width=9cm]{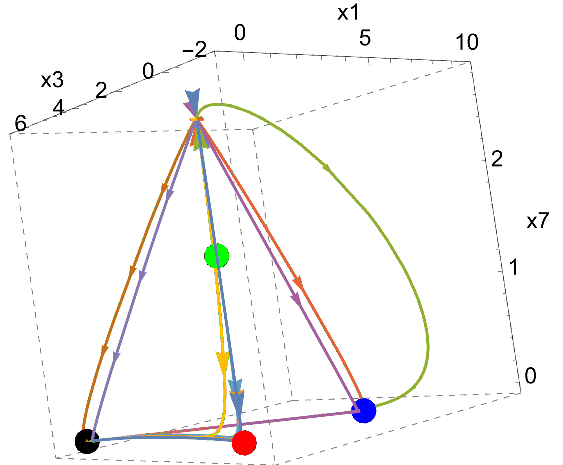}
		\caption{ 3D Phase portrait in $ x_1-x_3-x_7 $ plane for $ n =2, \xi = -1, \delta=1 $ and $m \in \mathbb{R} $ with coordinates of  red point $(0,0,0)$, blue point $(5,0,0)$, green point $(0,0,1.46)$, and black point $(0,7,0)$ .} 
\label{f5}
\end{figure}

\begin{figure}[h!]
\centering
		\includegraphics[height=9cm,width=9cm]{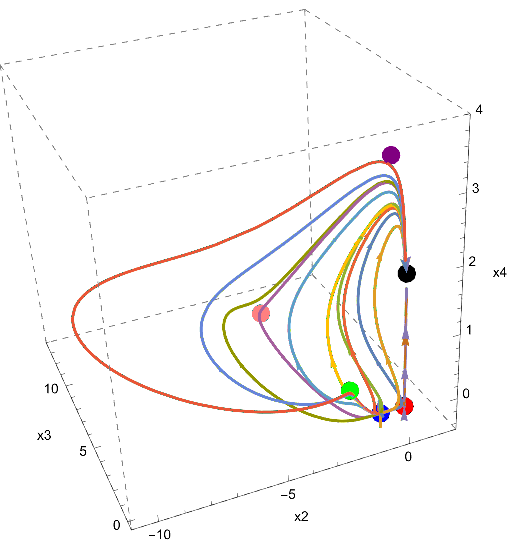}
		\caption{ 3D Phase portrait in $ x_2-x_3-x_4 $ plane for $ m =2, \delta=1, \xi \in \{-1, 0, 1\} $, and $n \in \mathbb{R} $  with coordinates of  red point $(0,0,0)$, blue point $\boldsymbol{(-1,0,0)}$, green point $(-\frac{9}{4},0,\frac{1}{2})$, black point $(0,0,2)$, pink point $(-\frac{7}{3},-\frac{28}{3},0)$ and purple point $(\frac{4}{19},-\frac{34}{19},\frac{10}{3})$. } 
\label{f6}
\end{figure}

\begin{figure}[h!]
\centering
		\includegraphics[height=9cm,width=9cm]{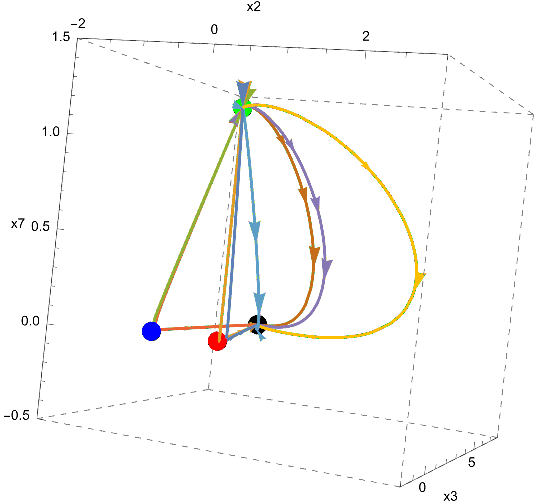}
  		\caption{ 3D Phase portrait in $x_2-x_3-x_7 $ plane for $ n =3, \delta=1/2, \xi \in \{-1, 0, 1\} $ and $m \in \mathbb{R} $ with coordinates of  red point $(0,0,0)$, blue point $(-1,0,0)$, green point $(0,0,\frac{57}{46})$, and black point $(-1,7,0)$.}
\label{f7}
\end{figure}

In Figures (\ref{f2} -\ref{f7}) , we plot the 3-dimensional projections of the 5-dimensional autonomous system (\ref{eq43a}-\ref{eq43e}). 

\clearpage

\section{Statefinder diagnostic analysis}\label{sec5}
The statefinder parameters involve the geometrical parameters such as the scale factor, Hubble parameters, and their derivatives. The statefinder parameters $ \{ r, s\}$ can be specified as $r=\frac{\dddot{a}}{aH^3}$, $s=\frac{r-1}{3(q-0.5)}$ \cite{Sahni2003}. Since these parameters involve the geometric parameters, one may use the dynamical variables (\ref{eq25}) to identify the behavior of statefinder parameters at the critical points in the cosmological dynamical system \cite{asgrg2024,as24a}. We use the dynamical variables (\ref{eq25}) with relation $dN=Hdt$ to convert the statefinder parameters into following form : 
\begin{align}
	r &= 3 - 9x_4 + 4x_4^2 - \frac{x_4 x_6}{m-1} \label{eqsf1}\\
	s &= \dfrac{2 - 9x_4 + 4x_4^2 - \frac{x_4 x_6}{m-1}}{3(\frac{1}{2} - x_4)} \label{eqsf2}
\end{align}
Whenever $ \{ r, s\}$ = $ \{ 1, 0\}$ in the $r-s$ plane is the fixed point of $\Lambda$ Cold Dark Matter ($\Lambda$CDM) model, whereas, for $ \{ r, s\}$ = $ \{ 1, 1\}$, it will resemble to the the Standard Cold Dark Matter (SCDM) model. For the varying dark energy in a model, the value of $r$ is not equal to $1$. Within the $r-s$ plane, the trajectories of the Chaplygin gas model and the quintessence model fall into distinct domains. In particular, trajectories associated with the quintessence model trace a path into regions characterized by  $ r < 1$  and $ s > 0$ . In contrast, the Chaplygin gas model trajectories will belong to the $r > 1$   $ s< 0$ region \cite{Sahni2003}.\\
The $r,s$ parameters are mainly used to discriminate between the dark energy behavior. The critical points F and I may have \( \{r, s\} = \{1, 0\} \), indicating a consistent alignment with the \(\Lambda\)CDM model. At the critical point, the parameters $\{r, s\}$  at the point $J$ are expressed as \[
\left\{ \frac{4m^4 - 6m^3 - 6m^2 + 7m + 3}{4m^4 - 12m^3 + 13m^2 - 6m + 1}, \frac{2(2 - m)}{3(2m^2 - 3m + 1)}\right\}.
\]
and $\{r, s\} = \{1, 0\}$,  for  $m=2$.\\  
\begin{figure}[h!]
\centering
		\includegraphics[height=8cm,width=12cm]{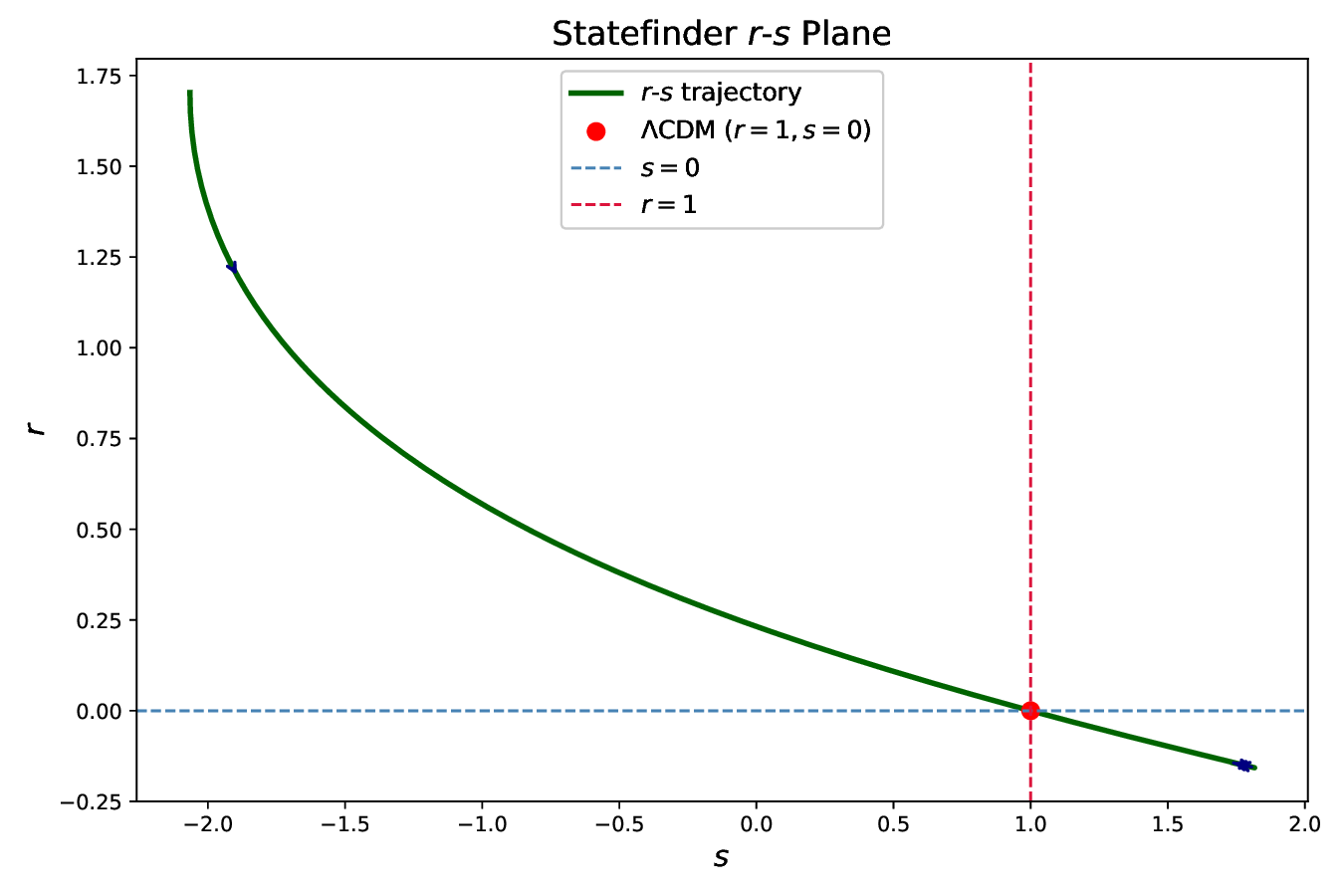}
		\caption{The $r-s$ plane diagram.} 
\label{f10}
\end{figure}
It indicates, based on the statefinder parameters at the critical point, that the model aligns with the $\Lambda$ CDM model at points F and I. This behavior is independent of the model parameters. Also, at point J, one may have the dynamical dark energy based on the parameter, $m$, which is the exponent of the Gauss-Bonnet term in $f(R,\mathcal{G})$ gravity form. This model may have quintessence-like characteristics based on point J. The $\Lambda$CDM dynamics having constant dark energy following $\omega_{eff}=-1$ is possible due to points F and I. The $r-s$ parameters depend on the variable $x_4$, and the behavior of these parameters has been given in Fig. \ref{f10}. The present model may unify the deceleration-acceleration transition of the past since the trajectory in the $r-s$ plane resembles the Chaplygin gas-like models. This trajectory crosses the $r=1,s=0$ meaning that the point J possessing quintessence behavior may act as the attractor in the model.
 \section{Cosmological Solutions, Observational Constraints and Cosmic Evolution } \label{sec6}
In this section, we use the cosmological solutions corresponding to the dynamical system. The critical point of the cosmological dynamical system corresponds to the cosmic phase of the universe evolution in model. The $f(R,\mathcal{G})$ model with interacting dark matter and scalar field possess different phases corresponding to points  $F$, $G$, $H$, $J$ and either of $A$ or $B/C/D/E$. The effective fluid corresponding to  point $A/B/C/D/E$ follows $\omega_{eff}=\frac{1}{3}$. In this effective phase, $\rho_r = \rho_{r0} a^{-4}$ which is primarily yielded by dynamical variables related to $f(R,\mathcal{G})$ function. The effective fluid corresponding to the points $F$ and $I$ will follow $\rho_\Lambda = \rho_{\Lambda0}$. And, the effective fluid corresponding to point $G$ will follow $\rho_{m1} =\rho_{m10} a^{\frac{-3(\xi+1)}{m}}$. For the point $H$, and $J$, the effective fluid will have $\rho_{m2} = \rho_{m20} a^{\frac{-6}{m}}$ and $\rho_{m3} = \rho_{m30} a^{\frac{-2(2-m)}{(2m-1)(m-1)}}$ respectively. These deductions are motivated with the fact that the effective EoS parameter during any cosmological phase may be related to the effective conservation equation $(\dot{\rho}+3H(1+ \omega_{eff})=0)$ in that phase. Using this analogy, we write the effective Hubble parameter of the model as \cite{Singh2025a}
\begin{equation}
    H^2={H_0}^2 \left[ \Omega_r a^{-4} + \Omega_\Lambda+ \Omega_{m1} a^{\frac{-3(\xi+1)}{m}}+ \Omega_{m2} a^{\frac{-6}{m}} + \Omega_{m3} a^{\frac{-2(2-m)}{(2m-1)(m-1)}}\right]
    \label{eq6a1}
\end{equation}

where $\Omega_r$, $\Omega_\Lambda$, $\Omega_{m1}$, $\Omega_{m2}$ and $\Omega_{m3}$ are the critical densities for the energy densities $\rho_r$, $\rho_\Lambda$, $\rho_{m1}$, $\rho_{m2}$ and, $\rho_{m3}$, respectively with $\Omega_r+\Omega_\Lambda+\Omega_{m1}+\Omega_{m2}+ \Omega_{m3}=1$. The scale factor $a$ and redshift $z$ is related by the relation $\frac{a_0}{a}=1+z$, where we take $a_0=1$, consistent with the standard assumption of observational cosmology. In the dynamical system section, we probed the cosmic dynamics from the late-times perspectives and here we aim to study the observational perspectives of model using the low redshift data. In the present observational study, $\Omega_{m3}$ is fixed at $\Omega_{m3}=0.001$. This choice is driven by the significant degeneracy of $\Omega_{m3}$ in the low-redshift Hubble data, making it difficult to constrain consistently alongside the other cosmological parameters using the combined dataset  consisting of cosmic chronometer (CC), Pantheon+SHOES type Ia supernovae (PS), CMB, and DESI DR2  baryon acoustic oscillation (BAO) \cite{Singh2025a}. \\
Moreover, In equation  (\ref{eq6a1}), it can be observed that the parameter $\xi$ appears only in the exponent of the $\Omega_{m1}$ term, i.e.,
$\Omega_{m1}\,a^{-3(\xi+1)/m}$. For $\xi=1$, one has $a^{-3(\xi+1)/m}=a^{-6/m}$, and therefore
the $\Omega_{m1}$ term has the same coefficient as the $\Omega_{m2}$ term. In this case, equation (\ref{eq6a1})
depends only on the combination
\begin{equation*}
\Omega_{m1}a^{-6/m}+\Omega_{m2}a^{-6/m} = (\Omega_{m1}+\Omega_{m2})a^{-6/m}
\equiv \Omega_{m12}\,a^{-6/m},
\end{equation*}
which implies that $\Omega_{m1}$ and $\Omega_{m2}$ are degenerate at the background level and only their sum $\Omega_{m12}$ is constrained by expansion data. Similarly, for $\xi=-1$, the $\Omega_{m1}$
contribution becomes constant ($a^{0}=1$) and can be absorbed into $\Omega_\Lambda$ via $\Omega_{\Lambda}^{'}=\Omega_\Lambda+\Omega_{m1}$. Therefore, the sign choice has no impact on the background dynamics and only introduces a parameter degeneracy. To avoid ambiguity, we fix $\xi=1$ and focus on the parameter combinations that are directly constrained by Eq.~(\ref{eq6a1}).
For $\xi =1$, the Eq. (\ref{eq6a1}) can be written as

\begin{equation}
H^2 = {H_0}^2 \left[ \Omega_r (1+z)^{4} + \Omega_\Lambda  + \Omega_{m12} (1+z)^{\frac{6}{m}} \right].
\label{6a2}
\end{equation}
Using Eq. (\ref{6a2}), we constrain the model parameters involved in $\omega_{eff}$ by using the observational datasets CC+PS+CMB+DESI DR2 BAO. As our interest lies in the late-time cosmological evolution, we use the standard relation $\Omega_{r0}=\Omega_{\gamma0}(1+0.2271N_{\nu})$ during the MCMC fitting process. Here, $\Omega_{\gamma0}=2.47\times10^{-5}h^{-2}$, $N_{\nu}\approx3.04$, and $h=H_0/(100\,\mathrm{km\,s^{-1}\,Mpc^{-1}})$ \cite{Aghanim2020}. These assumptions allow us to place observational constraints on the model parameters. \cite{Aghanim2020}.    
\subsection{The datasets and methodology}
\subsubsection{The cosmic chronometer data}

\par
\begin{figure}[h!]
       \centering
	\includegraphics[width=4in]{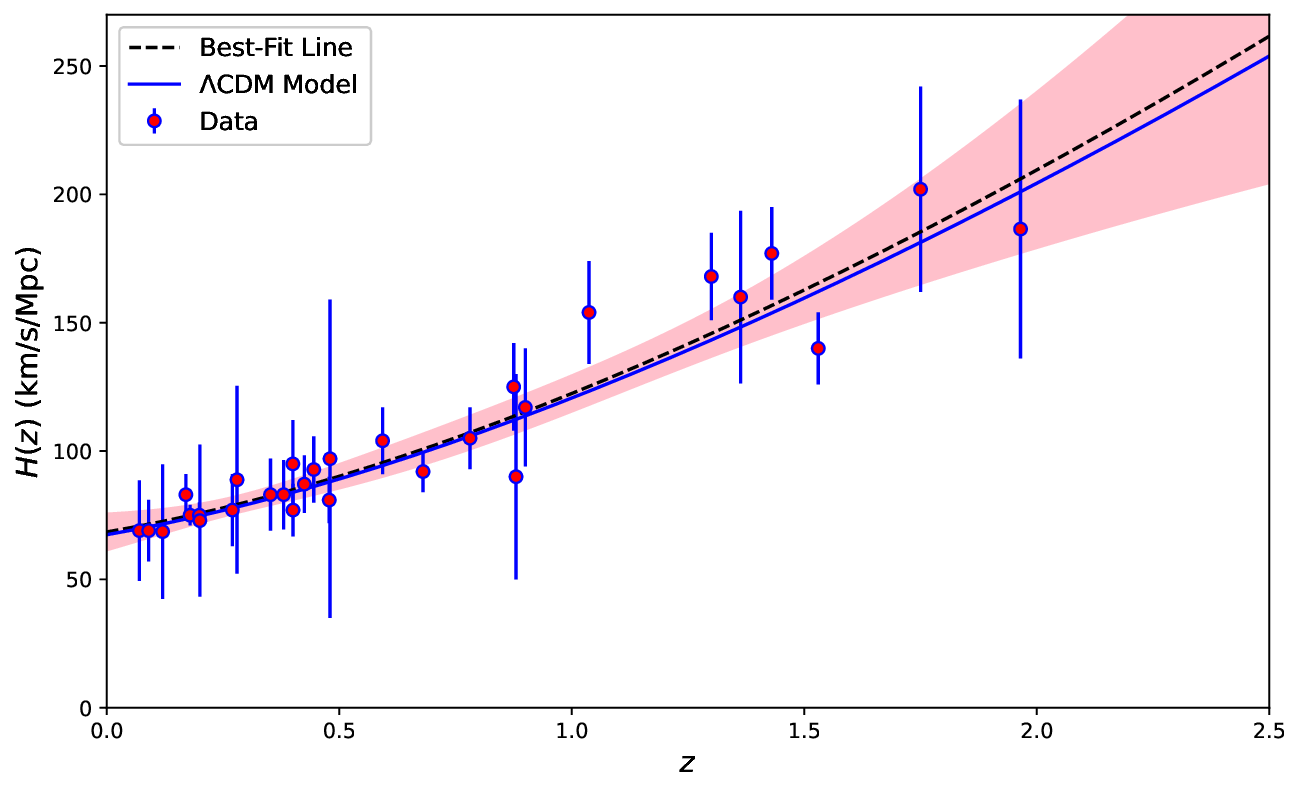}
	\caption{The best fit $H(z)$ curve with the cosmic chronometer (CC) data and its error bars
 \protect\label{fig6a1}}
\end{figure}

 The cosmic chronometer (CC) observations offer direct estimates of the Hubble expansion rate at different redshifts. The technique is based on the evolution of the scale factor $a$, which is related to the redshift $z$, leading to the expression $H(z)=-\frac{1}{1+z}\frac{dz}{dt}$ . These measurements provide valuable information about the expansion history of the universe \cite{jimenez2002}. In our analysis, we employ 31 cosmic chronometer data points consisting the redshift interval $0.07 \leq z \leq 1.965$ \cite{vagnozzi2021}. The likelihood associated with these observations is obtained using the $\chi^2$ function given by 

\begin{equation}
	\chi_{OHD}^2(\theta_p)=\sum_{i=1}^{31}\frac{\left[ H_{th}(\theta_p,z_i)-H_{obs}(z_i)\right]^2 }{\sigma_{H(z_i)}^2}
	\label{eq6a3}
\end{equation}
where $H_{th}(\theta_p,z_i)$ and $H_{obs}(z_i)$ are the theoretical and observed values of the Hubble parameter $H$, respectively.  and $\sigma_{H(z_i)}^2$ be the standard deviation for the observed value of each $H_{obs}(z_i)$. The best fit Hubble parameter curve in comparison with the $\Lambda$CDM model have been given in Figure \ref{fig6a1}.

\subsubsection{Pantheon + SHOES data}
To constrain the model parameters, we employ the Pantheon+SH0ES   supernovae type Ia (SNIa) dataset consisting of 1701 supernova measurements distributed over the redshift interval $0.00122 \leq z \leq 2.26137$. The Pantheon+SH0ES sample includes 1701 light-curve observations associated with 1550 spectroscopically identified Type Ia supernovae drawn from 18 different surveys \cite{brout2022, scolnic2022}. Throughout this work, we utilize the corrected apparent magnitude $m_b^{\rm corr}$ listed in the $m_b^{\rm corr}$ column of the dataset. Following Brout et al. \cite{brout2022}, we use the standard relation between the theoretical distance modulus and the luminosity distance. The distance modulus is defined as $\mu(z)=5\log_{10}\left(\frac{d_l(z)}{\rm Mpc}\right)+25$,
with the luminosity distance given by
$d_l(z)=c(1+z_{\rm hel})\int_{0}^{z_{\rm HD}}\frac{dz'}{H(z')}$. Here, $z_{\rm HD}$ and $z_{\rm hel}$ are taken from the `zHD' and `zhel' columns of the Pantheon+SH0ES dataset \cite{PantheonPlus} , respectively. The quantity $z_{\rm hel}$ denotes the heliocentric redshift of a supernova, whereas $z_{\rm HD}$ corresponds to the redshift corrected for the peculiar motions of galaxies and therefore represents the redshift in the cosmic rest frame \cite{brout2022, scolnic2022}. The parameter estimation is carried out by minimizing the $\chi^2$ function, which quantifies the agreement between the theoretical predictions and observational data. It is defined as \cite{brout2022, scolnic2022}
For the Pantheon+SH0ES sample, the likelihood is constructed using the $\chi^2$ function
\begin{equation}\label{60}
\chi^2_{\rm PP}=\Delta Q^{T}C^{-1}\Delta Q,
\end{equation}
where $C^{-1}$ is the inverse covariance matrix with dimensions $1701\times1701$ \cite{Pantheon} . The vector elements $Q_i$ are expressed as \cite{brout2022, scolnic2022}
\begin{equation}\label{61}
Q_i=
\begin{cases}
m_i - M - \mu_i^{\rm cep}, & \text{if } i \in \text{Cepheid hosts},\\
m_i-M-\mu_{\rm th}(z_i), & \text{otherwise}.
\end{cases}
\end{equation}
The quantity $\mu_i^{\rm cep}$ refers to the distance modulus of a Cepheid-calibrated host galaxy. These galaxies are identified through the `IS CALIBRATOR' column of the Pantheon+SH0ES data\cite{Pantheon}.

\begin{figure}[h!]
    \centering
    \includegraphics[width=4in]{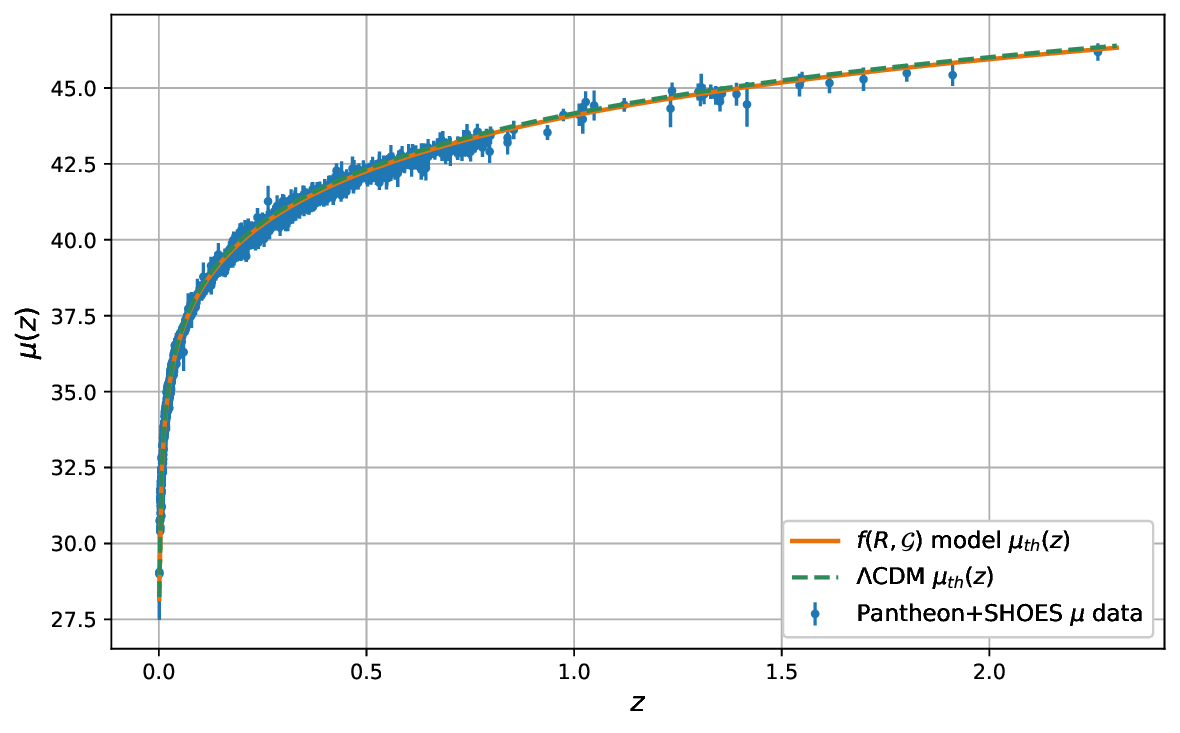}
   \caption{Comparison of the distance modulus $\mu(z)$ curve of the $f(R,\mathcal{G})$ model with the $\Lambda$CDM model and the Pantheon+SH0ES data.}
    \protect\label{fig6a2}
\end{figure} 

\subsubsection{CMB data}
To incorporate the CMB observations, we consider the compressed parameters $l_a$ and $R$, which are expressed as
\begin{equation}\label{62}
l_a=\pi \frac{D_M(z_*)}{r_s(z_*)},
\qquad
R=H_0\sqrt{\Omega_{m,0}}\int_0^{z_*}\frac{dz}{H(z)}.
\end{equation}
Here, $z_*$ represents the redshift of the last scattering surface, while $H_0=h\times100~{\rm km\,s^{-1}\,Mpc^{-1}}$. The likelihood is evaluated through
\begin{equation}\label{63}
\chi^2_{\rm cmb}=\Delta X\,C^{-1}_{\rm cmb}\,(\Delta X)^T.
\end{equation}
Here, $X=(R,l_a,\omega_b h^2)$, while $\Delta X=X_{\rm th}-X_{\rm obs}$ represents the difference between the theoretical predictions and the corresponding observational measurements. The observational values are taken from the Planck 2018 results \cite{Aghanim2020,chen2019}  and are given by $X_0 = (1.7428 \pm 0.0053,\ 301.406 \pm 0.090,\ 0.02259 \pm 0.00017)$.
\subsubsection{DESI-DR2-BAO data}
To further constrain the model, we make use of the BAO measurements from the DESI DR2 release \cite{abdul2025}. These data provide estimates of $D_V/r_d$, $D_M/r_d$, and $D_H/r_d$ across the redshift interval $0.295 \leq z \leq 2.33$, using tracers such as the Ly$\alpha$ forest, quasars, emission-line galaxies, bright galaxies, and luminous red galaxies \cite{abdul2025}. Here, $r_d=r_s(z_d)$ is the comoving sound horizon evaluated at the baryon drag epoch. The quantities $D_M$ and $D_H=c/H(z)$ denote the transverse comoving distance and the line-of-sight distance measure, respectively. The corresponding volume-averaged distance scale is defined through
$D_V=\left(zD_HD_M^2\right)^{1/3}$.\\
For the BAO analysis, we consider the DESI DR2 dataset \cite{abdul2025},  containing 13 observational data points of $D_V/r_d$, $D_M/r_d$, and $D_H/r_d$. The analysis also incorporates the corresponding cross-correlation coefficients to constrain the parameters of the decoupled power-law $f(R, \mathcal{G})$ model. It is worthwhile to mention that in this analysis, we treat $r_d$ as a free parameter \cite{pogosian2024,lin2021,vagnozzi2023} rather than fixing it using a prior derived from the Planck CMB observations.

\begin{figure}[h!]
 	\centerline{\includegraphics[width=5.1
    in]{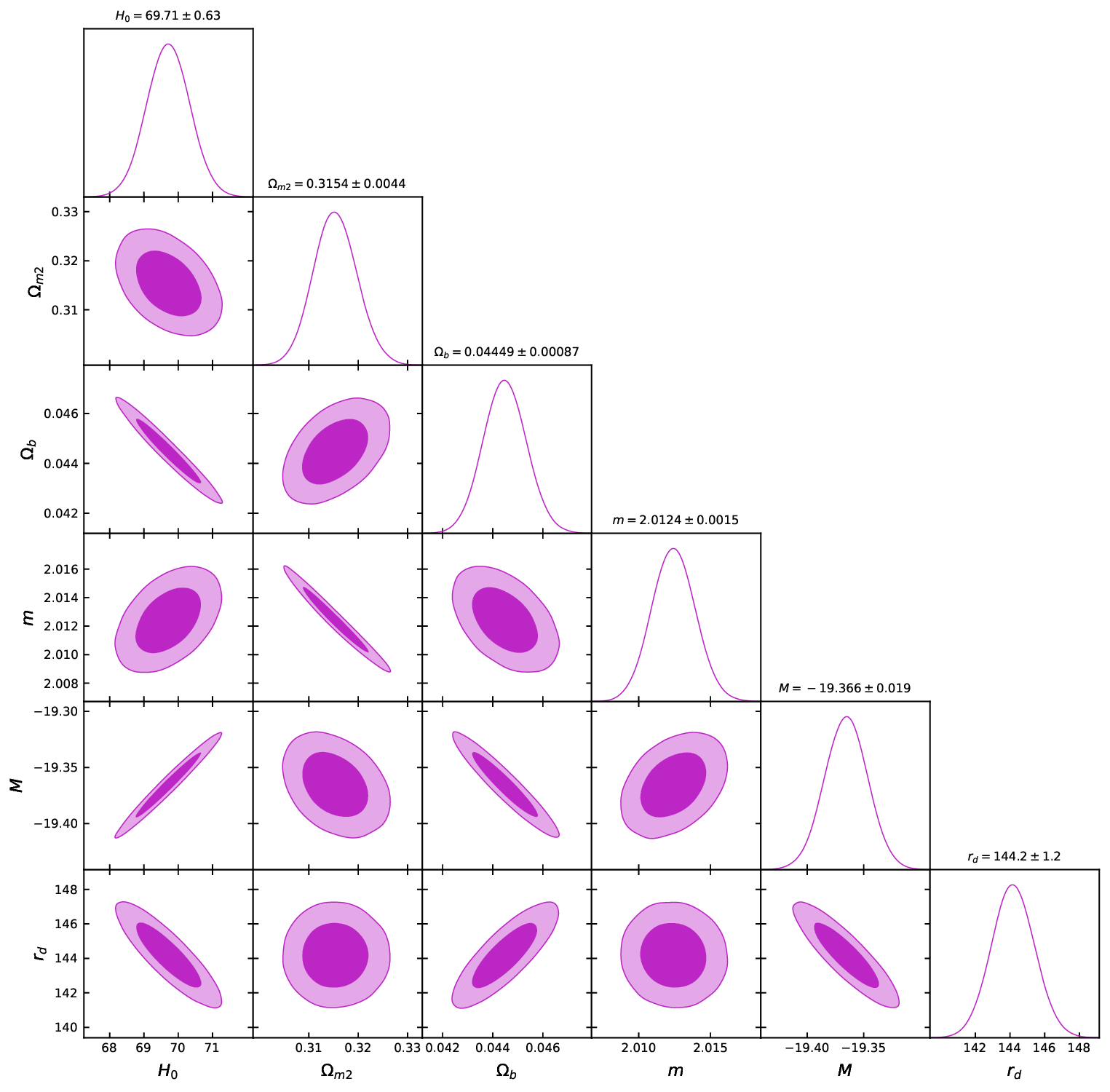}}\vspace*{8pt}
	\caption{The $1\sigma$--$2\sigma$ contour plots for the parameters $\{H_0, \Omega_{m2}, m, \Omega_b, M, r_d\}$ using the joint observational dataset.}
		\label{fig6a_c}
 \end{figure}

{\large Methodology} : The observational constraints are obtained from a joint dataset consisting of CC, CMB, Pantheon+SH0ES, and DESI DR2 BAO measurements. The corresponding total chi-square is written as
\begin{equation}
\chi_t^2=\chi_{cc}^2+\chi_{\rm cmb}^2+\chi_{\rm ps}^2+\chi_d^2.
\end{equation}
We then explore the parameter space using the MCMC sampler \texttt{emcee} \cite{ForemanMackey2013}. The analysis is performed with 60 walkers and 20,000 iterations for the considered model. The parameter set of the model is $(H_0,\Omega_{m2}, m, \Omega_b,M, r_d)$. The best-fit values of the model parameters are summarized in table \ref{t4}. The corresponding parameter space constrained by the combined CC+PS+DESI-DR2-BAO+CMB dataset is illustrated in figure \ref{fig6a_c}.

\subsection{Evolution of cosmological parameters}
\begin{figure}[h!]
    \centering
    \includegraphics[width=4in]{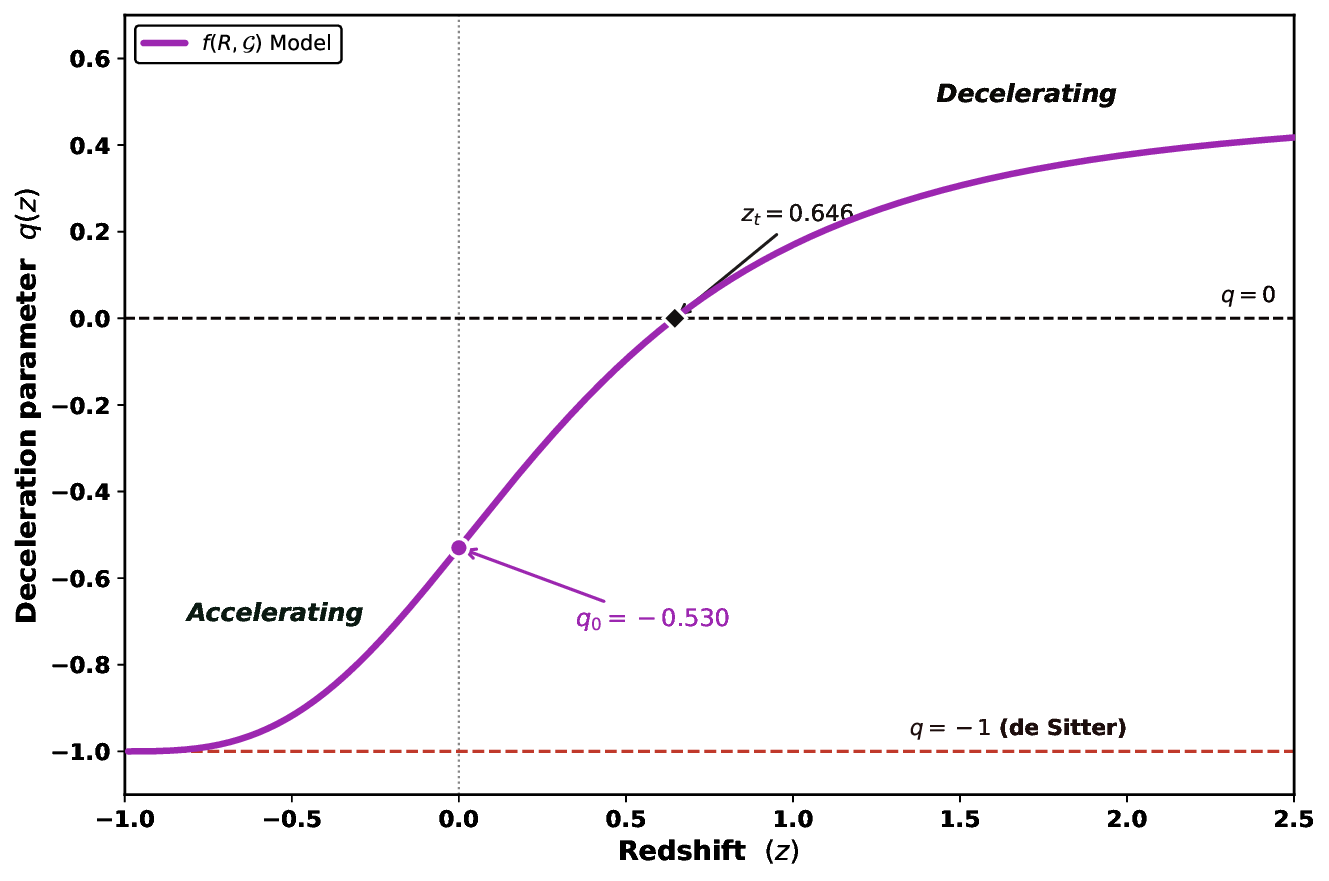}
    \caption{Evolution of the deceleration parameter $q(z)$ with redshift.}
    \protect\label{fig6a3}
\end{figure}

The cosmographic parameters provide a useful description of the expansion dynamics of the universe through the scale factor and its higher-order time derivatives. In this work, we focus on the deceleration, jerk, and snap parameters, denoted by $q$, $j$, and $S$, respectively. These quantities are defined as \cite{capozziello2019}
\begin{equation}\label{65}
q=-\frac{\ddot{a}}{aH^2}, \qquad
j=\frac{\dddot{a}}{aH^3}, \qquad
S=\frac{a^{(4)}}{aH^4}.
\end{equation}
using the relation
\begin{equation}\label{66}
\frac{d}{dt}=-(1+z)H\frac{d}{dz},
\end{equation}
they can be rewritten in terms of the redshift $z$ as

\begin{equation}\label{67}
q = -1 + \frac{(1+z)}{H}\frac{dH}{dz}, \qquad
j = (1+z)\frac{dq}{dz} + q(1+2q), \qquad
S = -(1+z)\frac{dj}{dz} - j(2+3q).
\end{equation}
In the de Sitter expansion regime of the $\Lambda$CDM model, the cosmographic parameters attain the values $q=-1$, $j=1$, and $S=1$. The sign of the deceleration parameter determines the expansion behavior, with $q<0$ indicating acceleration. In addition, the jerk parameter is a valuable diagnostic for identifying the transition between decelerating and accelerating phases of the universe.\\
The positive value of the Hubble constant $H_0$ in the $f(R,\mathcal{G})$ model signifies an expanding cosmological background. The observational constraints obtained here give $H_0 = 69.71 \pm 0.61~\mathrm{km\,s^{-1}\,Mpc^{-1}}$, which is separated from the Planck measurement \cite{Aghanim2020} by roughly $2.78\sigma$ and from the SH0ES measurement \cite{riess2022} by about $2.69\sigma$. This difference between the measurements is commonly referred to as the $H_0$ tension \cite{di2021}. The deceleration parameter is used to describe the change in the expansion rate over cosmic time. For the $f(R,\mathcal{G})$ model, the constrained present  value of the deceleration parameter $q_0=-0.530$ is negative, providing evidence for the present accelerated expansion of the universe. The variation of the deceleration parameter with redshift is shown in figure \ref{fig6a3}. The model predicts a transition from past deceleration to present acceleration at $z_t=0.646$ that has been shown in figure \ref{fig6a3}. In the distant future, the deceleration parameter converges to $q=-1$, pointing towards an alignment with the de Sitter state of the $\Lambda$CDM scenario.
\par The effective equation of state parameter $\omega_{\rm eff}$ provides insight into the properties of the cosmic fluid causing the expansion. We obtain $\omega_{{\rm eff},0}=-0.687$, which falls in the quintessence regime ($-1<\omega_{\rm eff}<-\tfrac{1}{3}$) and is consistent with the current accelerated expansion.
\begin{figure}[h!]
    \centering
    \includegraphics[width=4in]{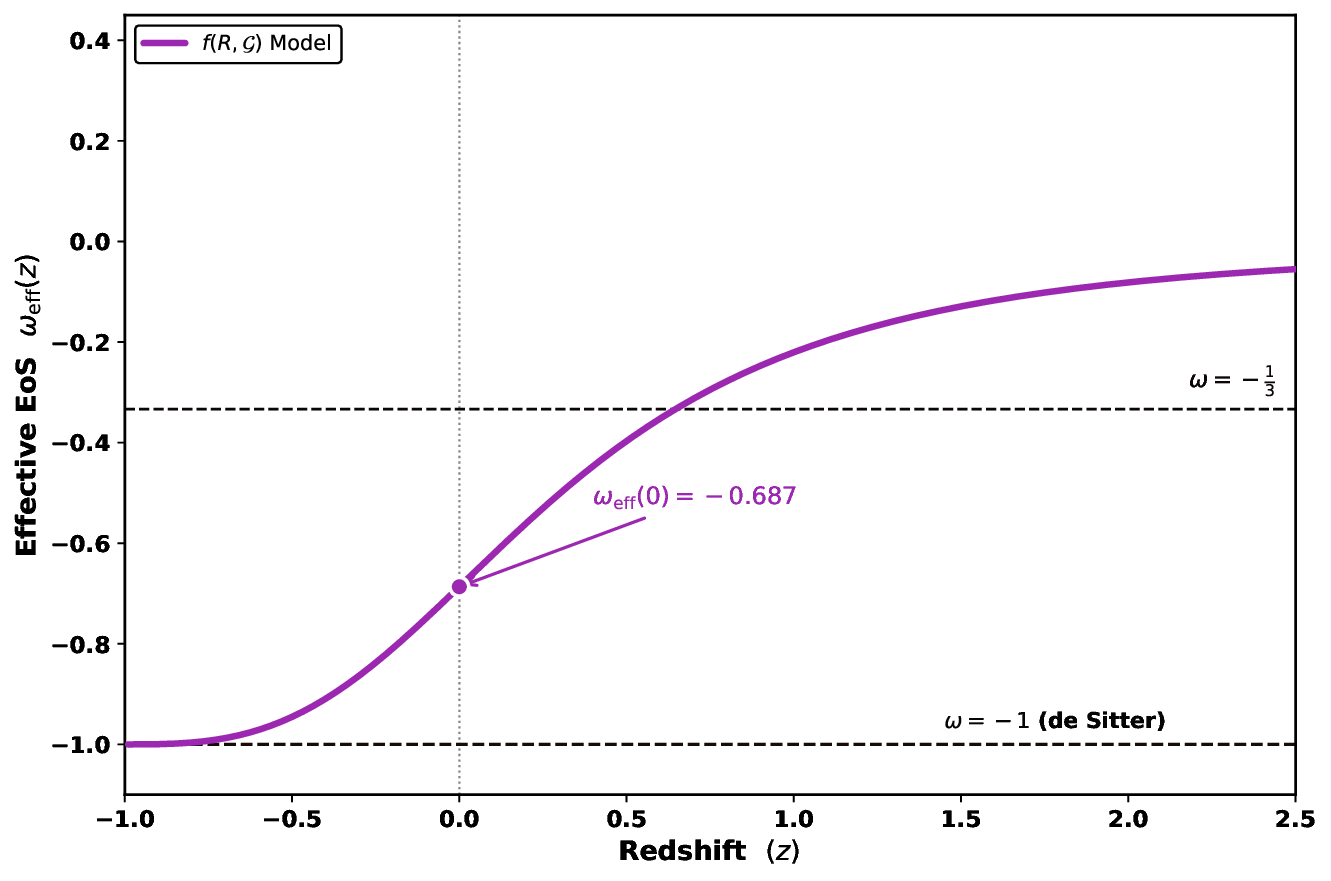}
    \caption{ Evolution of the effective equation of state parameter $\omega_{\rm eff}$ with redshift.}
    \label{fig6a4}
\end{figure} 
The evolution of $\omega_{\rm eff}$ with redshift is illustrated in figure \ref{fig6a4}. At higher redshifts, $\omega_{\rm eff}$ approaches the radiation and matter-dominated decelerating eras. As the universe evolves toward the far future ($z\rightarrow -1$), $\omega_{\rm eff}$ gradually approaches $-1$, indicating a de Sitter-like phase. As $\omega_{\rm eff}$ approaches $-1$, the $f(R,\mathcal{G})$ model gradually evolves toward a $\Lambda$CDM-like behavior in the late universe. Throughout the evolution, the phantom-divide line is not crossed.
\par
\begin{figure}[h!]
    \centering
    \includegraphics[width=4in]{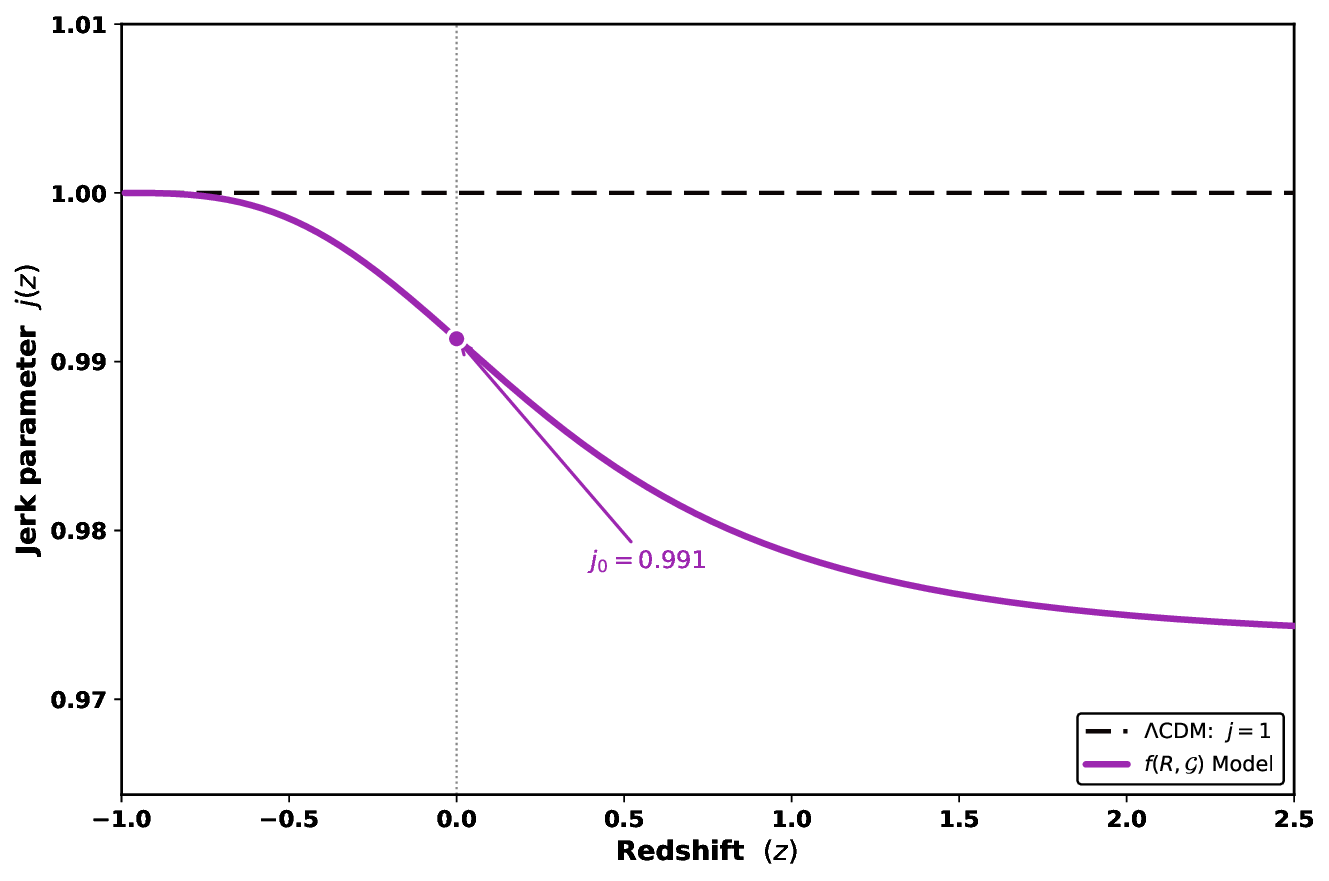}
    \caption{ Evolution of the jerk parameter $j(z)$ with redshift.}
    \protect\label{fig6a5}
\end{figure}

 At the present epoch, the jerk parameter takes the value $j_0=0.991$. The positive sign of $j_0$ is an indicator of a deceleration-acceleration transition, in agreement with the evolution of the deceleration parameter illustrated in figure \ref{fig6a5} and \ref{fig6a3}.
The jerk parameter stays nearly equal to unity over the entire redshift range and approaches $j=1$ at late times. This result indicates that the curvature-induced dynamics of the $f(R,\mathcal{G})$ model yield an expansion history very close to the standard  $\Lambda$CDM model.
\par
\begin{figure}[h!]
    \centering
    \includegraphics[width=4in]{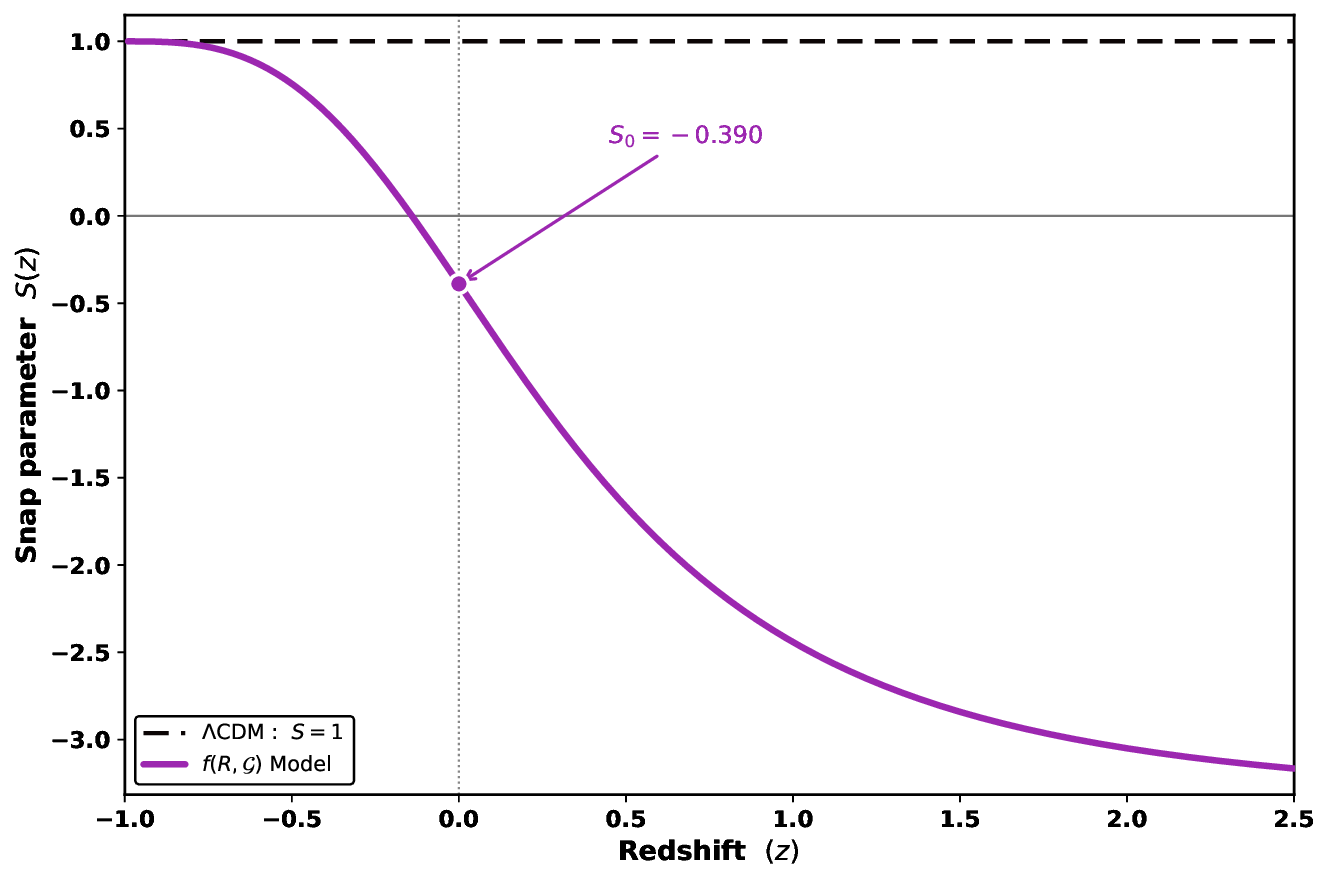}
    \caption{Evolution of the snap parameter $S(z)$ with redshift.}
    \label{fig6a6}
\end{figure}

 The snap parameter is constrained by the value $S_0=-0.390$ at the present epoch. Since the de Sitter expansion of the $\Lambda$CDM scenario corresponds to $S=1$, the current universe is still far from this asymptotic state. The redshift evolution of the snap parameter is depicted in figure \ref{fig6a6}. The parameter remains negative during the earlier decelerating stages of cosmic evolution and gradually converges to the $\Lambda$CDM limit, $S=1$, in the distant future.
\par
\begin{figure}[h!]
    \centering
    \includegraphics[width=4in]{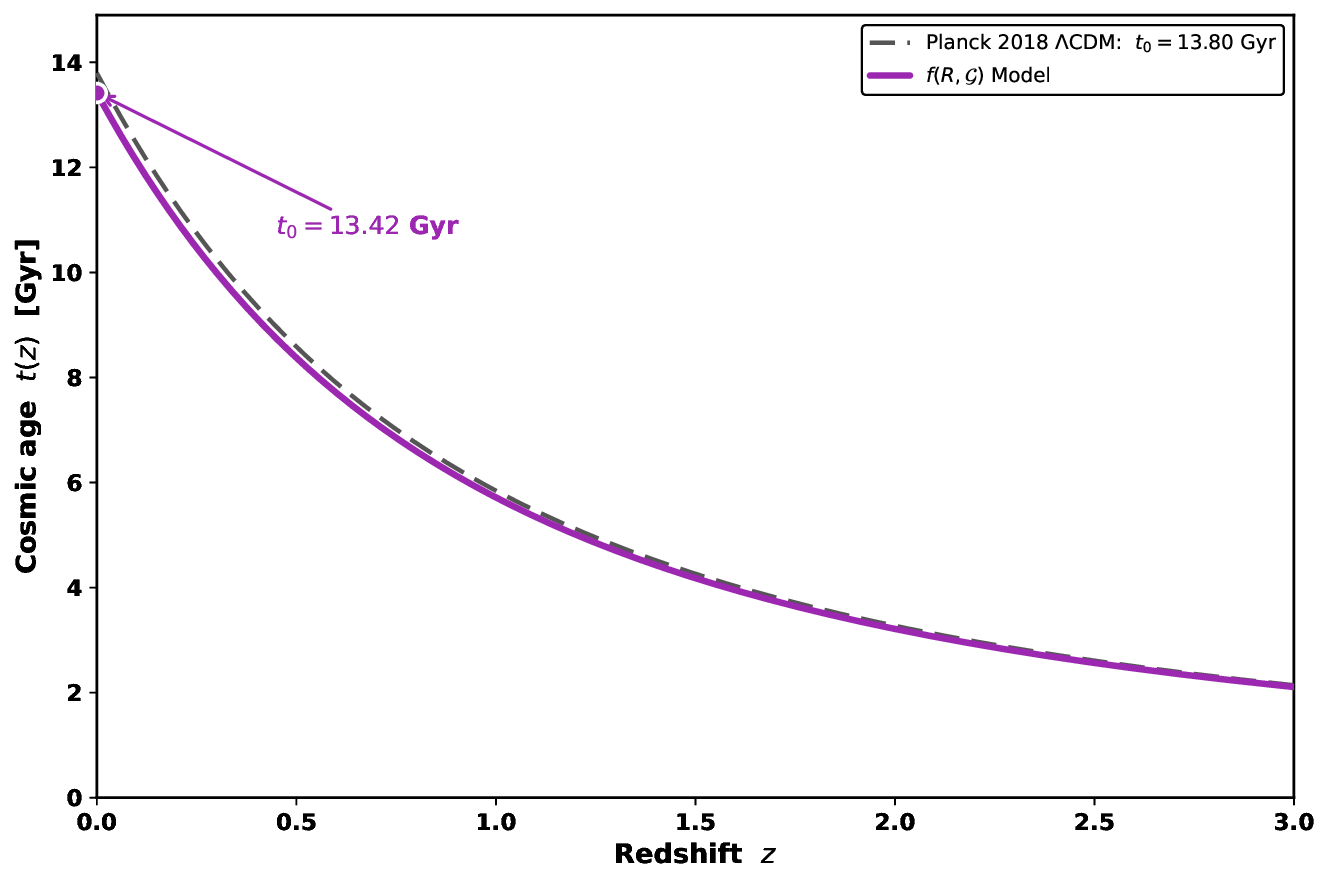}
    \caption{Evolution of the cosmic age t(z) with redshift.}
    \protect\label{fig6a7}
\end{figure}

 The age of the universe in the $f(R,\mathcal{G})$ model is obtained from $t(z)=\int_{z}^{\infty}\frac{dz'}{(1+z')H(z')}$,
which yields a present age of $t_0=13.42$ Gyr which is given in figure \ref{fig6a7}. This estimate is broadly consistent with the Planck 2018 $\Lambda$CDM value of $13.80$ Gyr and lies within the range supported by observational Hubble data.
The consistency of the estimated cosmographic parameters and cosmic age with the corresponding $\Lambda$CDM predictions demonstrates that the $f(R,\mathcal{G})$ model provides a viable description of the late-time universe.

\begin{table}[htbp]
\centering
\caption{The median values with $1\sigma$ errors for the Decoupled Power-Law
$f(R,\mathcal{G})$ model, where the joint dataset is composed of CC,
DESI-DR2-BAO, CMB and Pantheon+SH0ES. The unit of $H_0$ is km/s/Mpc.}
\label{t4}
\begin{tabular}{lc}
\hline\hline
Model Parameter & $f(R,\mathcal{G})$ model \\
\hline
$H_0$        & $69.71 \pm 0.63$ \\
$\Omega_m$   & $0.3154 \pm 0.0044$ \\
$\Omega_b$   & $0.04449 \pm 0.00087$ \\
$m$          & $2.0124 \pm 0.0015$ \\
$M$          & $-19.366 \pm 0.019$ \\
$r_d$        & $144.2 \pm 1.2$ \\
\hline
\multicolumn{2}{l}{Cosmological parameters} \\
$q_0$         & $-0.530$ \\
$j_0$         & $0.991$ \\
$S_0$         & $-0.390$ \\
$\omega_{\mathrm{eff},0}$ & $-0.687$ \\
$t_0$         & $13.42$ \\
\hline
\multicolumn{2}{l}{Information criterion} \\
$\chi^2_{\min}$ & $1651.25$ \\

\hline\hline
\end{tabular}
\end{table}

\clearpage

\section{Conclusions}
\label{section6}
This study contributes to the understanding of the later stages of cosmic expansion by exploring the modified $ f(R, \mathcal{G}) $ gravity theory with scalar field interacting with matter. We proceed with a nonlinear Decoupled Power-law $ f(R, \mathcal{G}) $ model admitted by Noether symmetry analysis given by $ f(R, \mathcal{G}) = \alpha R^m +  \beta \mathcal{G}^n  $, with parameters $ m, n, \alpha, $ and $ \beta $ having no constraints. In the cosmological  $ f(R, \mathcal{G}) $ model, we investigated the mathematical formulation of equations of motion particular to a flat FLRW universe and found an autonomous system characterized by a system of equations (\ref{eq43a}-\ref{eq43e}). We have also shown the eigenvalues, critical points, and their presence in the given model through a detailed dynamical systems analysis using the equation of state.\\
The autonomous system derived for the chosen \( f(R, G) \) model yields ten critical points. Either of the points A, B, C, D, and E will have the radiation-dominated phase in the effective sense. The non-linearity associated with the $f(R,\mathcal{G})$ function form leads to the existence of these radiation-dominated points. These points will have a non-zero contribution from the dynamical variable, and their stability character will depend on different model parameters, but the phase governed by these points will have $q=1$, $\omega_{eff}=\frac{1}{3}$. The points F and I would correspond to the de Sitter expansion phase and are primarily driven by dark energy having $q=-1$ and $\omega_{eff}=-1$. However, the dynamical expansion phases of points $F$ will have the contribution from the potential term and the Ricci scalar. On the other hand, at the point $I$, the de Sitter phase will be due to the non-zero contribution from the $R$ and $\mathcal{G}$ term in this modified gravity model with  an interacting scalar field. The points G and H stand out due to their dependence on the parameter \( m \). For instance, when \( m = \xi + 1 \), point G aligns with a matter-dominated era, and point H similarly represents a matter-dominated universe when \( m = 2 \). In contrast, setting \( \xi = -1 \) results in a dark energy-dominated scenario at point G, and at point H, the universe is again influenced by dynamical dark energy for \( m = \frac{3}{2} \). Finally, point J indicates a dark energy-dominated phase when \( m = -1 \), and it returns to a matter-dominated regime for \( m = \frac{7 \pm \sqrt{73}}{12} \). The dynamical behavior of the system has been illustrated using 3D phase-plane diagrams in Figures  \ref{f2}, \ref{f3}, \ref{f4},  \ref{f5},  \ref{f6}, and \ref{f7}.\\
The dynamics of the universe's evolution are further explored through the investigation of the scale factor $ a $ and the deceleration parameter $ q$. The phase space analysis indicates two potential sequences of cosmic evolution based upon certain limitations on the parameters $ m$, $n, \xi,$ and $\delta$. The radiation, matter and dark energy-dominated era may be explained in this model based on the critical points. The trajectories depict physically possible cosmic pathways, wherein the universe evolves from an initial unstable condition, undergoes a phase of matter dominance, and ultimately reaches a stable acceleration phase. For the explanation of the radiation phase, any of the points A, B, C, D, and E will be suitable candidates. On the other hand, points F, I, and J represent the impact of dark energy. Likewise, point G may either have the matter, radiation, and the dark energy when the values of the parameter $ m= \xi+1, \frac{3(1+\xi)}{4},$ and $ \xi = -1 $ respectively. In the same manner, the point H is dominated by matter and radiation for $m=2$ and $m=3/2$ respectively. The matter-dominated era in this model will be explained either by point G or H. The non-zero value of $x_1$ for point G makes it a more suitable candidate as compared to point H.\\
Apart from scrutinizing the phase space of dynamical variables, we probe the nature of state-finder parameters to ascertain the kind of dark energy in the present model. Points F and I correspond to the $\Lambda$CDM point, while point J may have dark energy with varying energy.  The resulting trajectories in the $(r, s)$ plane, as illustrated in Figure \ref{f10}, demonstrate that the system initially deviates from, but ultimately evolves toward, the $\Lambda$CDM fixed point, thereby offering valuable insights into the model’s late-time cosmological consistency.
\\
The present modified gravity based interacting dark energy model offers an interesting class of asymptotic solutions for radiation, matter and dark energy-on dominated eras of the universe. Based on this promising phenomenology of interacting dark energy dynamics in $f(R,\mathcal{G})$ gravity, the precise statistical analysis based on constraints subjected to the observational data may describe the cosmic evolution in a more descriptive way. In addition, the growth of matter perturbations will further describe the compatibility of model with the early universe dynamics and  the structure formation era. We are working to study these aspects in the very near future.
\par  The effective equation of the state parameter corresponding to the critical points is utilized to write the effective Hubble parameter of the model. To assess the observational consistency of the model, we perform an MCMC analysis using a joint dataset consisting of cosmic chronometer, Pantheon+SH0ES, CMB, and DESI DR2 BAO measurements. A comprehensive summary of the constrained model parameters, derived cosmological parameters, and the corresponding minimum chi-square value is presented in table \ref{t4}. As shown in figures .\ref{fig6a1} and \ref{fig6a2}, the reconstructed Hubble parameter and distance modulus closely follow the observational data and are consistent with the $\Lambda$CDM model. The corresponding confidence contours are presented in figure \ref{fig6a_c}. The analysis yields $H_0 = 69.71 \pm 0.61~{\rm km\,s^{-1}\,Mpc^{-1}}$, which is approximately $2.78\sigma$ away from the Planck estimate and $2.69\sigma$ away from the SH0ES value, indicating the well-known $H_0$ tension. The present value of the deceleration parameter, $q_0=-0.530$, confirms that the universe is currently undergoing accelerated expansion. The transition from deceleration to acceleration occurs at $z_t=0.646$, as shown in figure \ref{fig6a3}. Furthermore, the effective equation of state parameter, $\omega_{{\rm eff},0}=-0.687$, lies within the quintessence regime. The gradual evolution toward $\omega_{\rm eff}=-1$ in the far future suggests a late-time approach to $\Lambda$CDM-like behavior without crossing the phantom-divide line (figure \ref{fig6a4}). The cosmographic analysis also points toward a cosmological evolution close to that of $\Lambda$CDM. As shown in figure \ref{fig6a5}, the present jerk parameter is $j_0=0.991$, remaining nearly equal to unity, while the snap parameter, $S_0=-0.390$ (figure \ref{fig6a6}), tends toward the de Sitter limit in the asymptotic future. The age of the universe is estimated to be $t_0=13.42$ Gyr (figure \ref{fig6a7}), which is close to the Planck 2018 $\Lambda$CDM value of $13.80$ Gyr and further supports the observational viability of the model.\\
In summary, the dynamical system analysis together with the observational constraints supports a quintessence-like late-time acceleration, with the cosmic evolution gradually converging to the $\Lambda$CDM limit. The combined dynamical and observational evidence indicates that the interacting $f(R,\mathcal{G})$ model captures the essential features of the universe's evolution. These encouraging results motivate further examination of the model through perturbation analyses and structure formation studies.

\section*{\textbf{Acknowledgement}}
Shivani, acknowledges CSIR-UGC, New Delhi, for the financial aid provided under the CSIR-UGC(JRF) scheme with the UGC-Ref.No.: 1332/(CSIR-UGC NET JUNE 2019). RC thanks SERB, New Delhi, for financial assistance through project No. CRG/2023/004560 (P-07/1328).

\section*{Data Availability}

There are no new data associated with this article.
\medskip

 \bibliographystyle{ieeetr}

\bibliography{Shivani1}

\end{document}